\documentclass[aps,prb,onecolumn,superscriptaddress,10pt]{revtex4-2}

\usepackage{amsmath,amssymb}
\usepackage{graphicx}
\usepackage{overpic}
\usepackage{subfigure}

\usepackage{hyperref}
\usepackage[margin=1.30in]{geometry}

\renewcommand{\Im}{\mathrm{Im}}

\begin{document}

\title{Disorder-Assisted Adiabaticity in Correlated Many-Particle Systems}

% \author{Aly Abuelmaged}
% \affiliation{Department of Physics, University at Buffalo SUNY, Buffalo, New York 14260, USA}

\author{Shang-Jie Liou}
\affiliation{Department of Physics, University at Buffalo SUNY, Buffalo, New York 14260, USA}

\author{Herbert F.~Fotso}
\affiliation{Department of Physics, University at Buffalo SUNY, Buffalo, New York 14260, USA}

\begin{abstract}
\noindent We investigate how disorder affects adiabaticity in an interacting quantum system by assessing its effect on the state of the system after an interaction modulation, or interaction ``pulse" ,whereby the interaction is changed from zero to a maximum value and then back to zero following a given time profile. We find that, independently of the disorder strength and pulse shapes (rectangular, triangular, and Gaussian), the pulse duration is negatively correlated with the change in total energy in the system. That is, the longer duration reduces the change in total energy for each protocol. Most importantly, across different considered pulse shapes, we find a robust negative correlation between the disorder strength and the change in total energy across the interaction pulse. Namely, increasing the disorder strength systematically suppresses the residual energy added to the system after the interaction pulse, indicating a more adiabatic response. These two effects, disorder-induced and duration-induced adiabaticity, are consistently observed across all three pulse shapes. Among the protocols, the triangular pulse yields the smallest change in total energy in the system over comparable conditions, demonstrating the most adiabatic response. In addition to the energy analysis, we also examine how disorder modifies the effective temperature change across the interaction pulse, to further establish a quantitative relation between disorder and the thermal response. Altogether, our results identify disorder as a key factor in both the energy and the temperature variation over the time-modulation of the interaction.
\end{abstract}


\maketitle

%%%%%%%%%%%%%%%%%%%%%%%%%%%%%%%%%%%%%%%%%%

\section{Introduction}
\label{sec:Introduction}
\noindent Adiabaticity is a fundamental concept with applications in a variety of fields extending from thermodynamics\cite{ag1,ag2, ag3, ag4}, to quantum information processing\cite{qc1, qc2, qc3}, to quantum state preparation\cite{sp1, sp2, sp3, sp4}, and, more broadly, to many-body physics\cite{mbd1, mbd2, mbd3} . For quantum systems, adiabaticity is often related to the rate of change of the Hamiltonian and powers of the minimum energy separation between consecutive energy levels of the Hamiltonian\cite{ag2, ad2, ad3, ad4, ad5, ad6}. It has been extensively investigated for the preparation of quantum states of strongly correlated systems in optical lattices\cite{sp1, sp2, sp3}. A general challenge for correlated many-particle systems is the difficulty of computing the energy gaps for large many-body systems. Also, given that disorder is typically a feature rather than the exception in realistic systems, it is important to understand its effect on the adiabaticity of many-particle systems.

In the present paper, we aim analyze the dynamics of a many-body system across a modulation of the interaction on a disordered system. Namely, we consider a system described by the Anderson-Hubbard model that features itinerant electrons on a lattice, a random site energy, and a Coulomb interaction for doubly occupied sites. We perform a time modulation of the interaction, or an interaction "pulse", whereby the interaction starts off at zero and is increased to a maximum value before then being switched off again, according to a given pulse shape (rectangular, triangular, and Gaussian). We use our recently introduced Nonequilibrium DMFT+CPA\cite{NeqDMFT_CPA1, NeqDMFT_CPABinaryDisorder, NeqDMFT_CPAThermalization, NeqDMFT_CPA2} method that combines the nonequilibrium extensions of the dynamical mean field approximation (DMFT)\cite{NeqDMFT1, NeqDMFT2, NeqDMFT3, NeqDMFT4, NeqDMFT5, NeqDMFT6, NeqDMFT7, NeqDMFT8} and that of the coherent potential approximation (CPA)\cite{NeqCPA1, NeqCPA2, NeqCPA3, NeqCPA4, NeqCPA5, NeqCPA6}, to appropriately treat the nonequilibrium dynamics of our interacting and disordered system. We probe adiabaticity through the residual total energy in the system after the interaction pulse.

Through these solutions, we find that independently of the disorder strength and pulse shape, longer pulse durations reduce the change in total energy in the system, as would be generally expected. Most importantly, across the different pulse shapes considered, we find a robust negative correlation between disorder strength and the residual energy added to the system after the interaction modulation. Namely, increasing the disorder strength systematically suppresses the change in total energy after the interaction is returned to zero, indicating a more adiabatic response of the system to the interaction pulse. 

These two effects, disorder-induced and duration-induced adiabaticity, are consistently observed across all three protocol shapes. Among the pulse shapes considered, the triangular protocol displays the smallest change in total energy in the system under comparable conditions, demonstrating the most adiabatic response. In addition to the energy analysis, to establish a quantitative relation between disorder and the thermal response, we examine how disorder affects the effective temperature change during the interaction modulation process. Consistently with the change in total energy, the variation in temperature across the interaction pulse is reduced by both the pulse duration and the disorder strength. Taken altogether, our results identify disorder, protocol duration, and protocol shape as key factors governing both energy and temperature evolution during interaction modulation i.e. as control parameters for more adiabatic evolution.

The rest of the paper is structured as follows. In section~\ref{sec:ModelMethods}, we discuss our model and its solution with our nonequilibrium DMFT+CPA approach. In section~\ref{sec:Results}, we present our results before concluding with section~\ref{sec:Conclusion}.

%%%%%%%%%%%%%%%%%%%%%%%%%%%%%%%%%%%%%%%%%%
\section{Model and Methods}
\label{sec:ModelMethods}
\subsection{Model}
\noindent We consider a disordered correlated electronic system described by the single-band Anderson--Hubbard model, initially in equilibrium at temperature $T = 1/\beta$. The Hamiltonian is
\begin{equation}
H = -\sum_{\langle i j \rangle, \sigma} t_{ij}
\left(c^{\dagger}_{i\sigma} c_{j\sigma} + \text{H.c.}\right)
+ \sum_i U(t)\, n_{i\uparrow} n_{i\downarrow} 
+ \sum_{i\sigma} (V_i - \mu)\, n_{i\sigma}.
\label{eq:Hamiltonian}
\end{equation}

Here, $t_{ij} = t_{\text{hop}}$ is the hopping amplitude between nearest-neighbor sites denoted by $\langle i j \rangle$, $U(t)$ is the on-site Coulomb interaction at time $t$, and $V_i$ is the random on-site disorder potential. The operators $c^{\dagger}_{i\sigma}$ ($c_{i\sigma}$) create (annihilate) an electron of spin $\sigma = \uparrow, \downarrow$ at site $i$, and $n_{i\sigma} = c^{\dagger}_{i\sigma} c_{i\sigma}$ is the corresponding number operator. The chemical potential $\mu$ is set to $U/2$ to ensure half-filling. We work in units where $\hbar = e = c = 1$.
The disorder potential $V_i$ follows a uniform box probability distribution between $-W$ and $+W$, 
$
P(V_i) = \frac{1}{2W}\,\Theta(W - |V_i|),
$
where $W$ characterizes the strength of the disorder. 
The problem is solved on the Bethe lattice in the limit of infinite coordination number $z \to \infty$, for which the hopping amplitude is rescaled as $t_{\text{hop}} = t^{*}/\sqrt{z}$. %Since the bandwidth of a clean non-interacting system is $4t^{*}$
 We choose $t^{*} = 0.25$ and we use  the bandwidth $4t^{*} = 1$ as the unit of energy and time.

Under equilibrium conditions, the interaction \(U(t)\) remains constant. 
In contrast, for the nonequilibrium scenarios of interest in the present paper, \(U(t)\) varies in time according to a given profile. In this work, we examine three distinct "pulse" shapes --- rectangular, triangular, and Gaussian --- each characterized by the maximum amplitude $U_{max}$ and designed such that $U(t) \rightarrow 0$ at sufficiently early and late times. The disorder potential $\{V_i\}$ remains constant throughout the evolution.

In the rectangular pulse case, the interaction is abruptly switched on from $U = 0$ to a finite value $U_{\text{max}}$ for a fixed duration $T_p$, and then switched off again, following $U(t) = 0$ for $t < 0$, $U(t) = U_{\text{max}}$ for $0 \le t < T_p$, and $U(t) = 0$ for $t \ge T_p$. This form allows us to analyze the response of the system to a sudden and finite-duration interaction pulse. On the other hand, the triangular pulse rises linearly from zero at $t = 0$ to $U_{\text{max}}$ at time $t = t_0$ and symmetrically decays back to zero, with a total pulse width $T_p$, such that $U(t) = 0$ for $|t - t_0| > T_p/2$ and $U(t) = U_{\text{max}} \!\left(1 - |t - t_0|/(T_p/2)\right)$ for $|t - t_0| \le T_p/2$. The triangular pulse profile features smoother ramping and decay than the rectangular pulse. Finally, the Gaussian pulse varies continuously according to $U(t) = U_{\text{max}} \exp[-(t - t_0)^2/(2\sigma^2)]$, where $t_0$ is the pulse center (with $U(t_0) = U_{\text{max}}$) and $\sigma$ denotes the standard deviation of the pulse. $\sigma$ is chosen such that $U(0) = 0.01\,U_{\text{max}}$, giving $\sigma = \sqrt{-t_0^2 / [2\ln(0.01)]}$; also, the pulse width is determined by $T_p=2t_0$. These three time-dependent profiles of $U(t)$ represent different ways of modulating the interaction, allowing us to investigate how the shape of the interaction strength influences the disordered system’s ability to respond adiabatically. Each type of pulse starts and ends at zero interaction, meaning that the system returns to its initial non-interacting configuration. For these different pulse shapes, we will evaluate the residual energy, change in total energy, in the system after the pulse. We will also assess the temperature change due to the pulse.

%By applying these pulses to the system, we can examine how the presence of disorder alters the system’s response to the interaction modulation, thereby revealing whether disorder enhances or suppresses quantum adiabaticity during the dynamical process.

\subsection{Nonequilibrium DMFT+CPA}

\begin{figure}[htbp]
\centering
\includegraphics[width=\columnwidth]{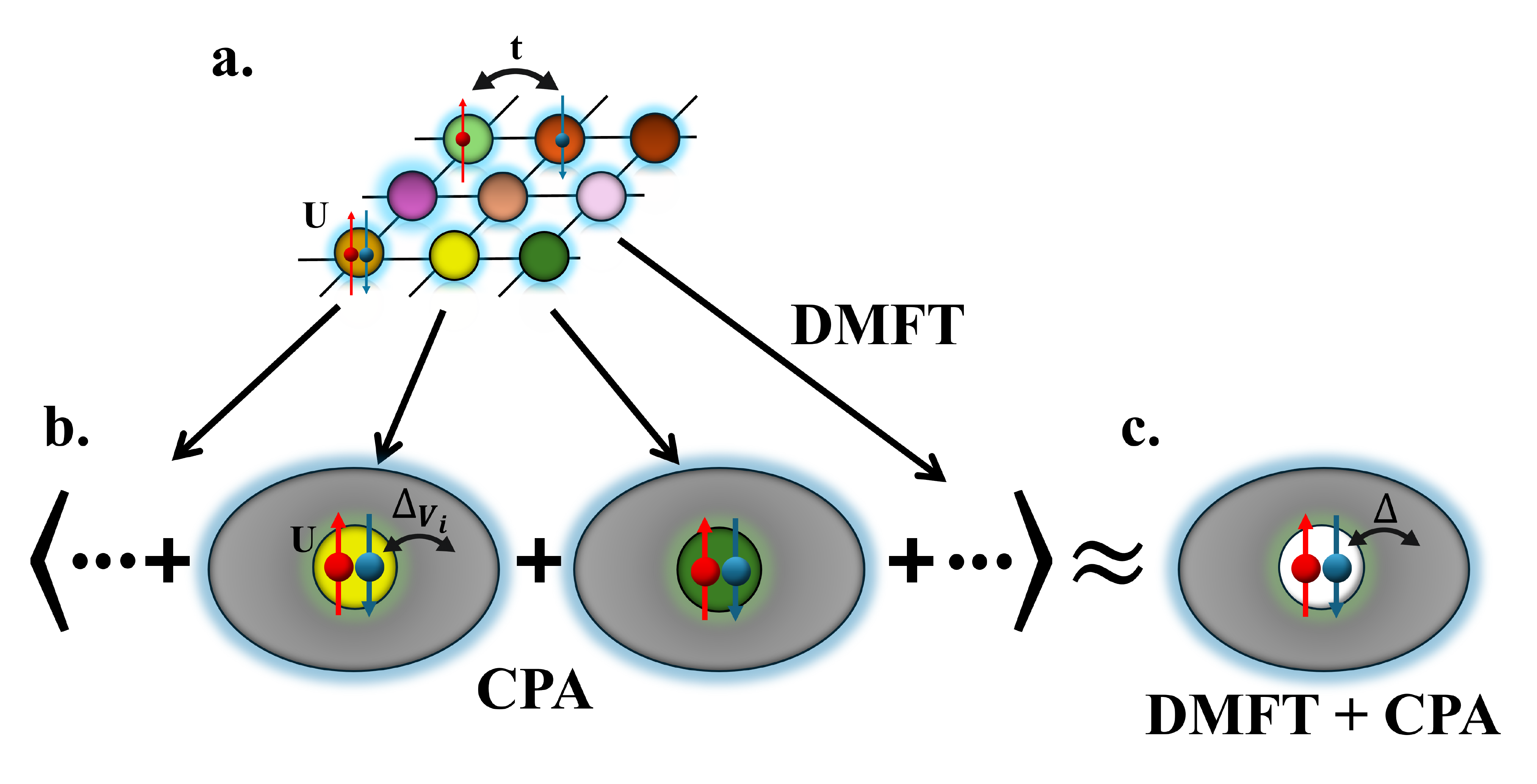}
\caption{Schematic illustration of the DMFT+CPA solution: 
(\textbf{a}) Anderson-Hubbard model, electrons can hop between the nearest neighboring lattice sites $i$ and $j$ with hopping amplitude $t_{ij}$, experience an on-site interaction energy $U$ for doubly occupied sites , and are subject to random disorder potential $V_{i}$ (shown as different colors in the figure).
(\textbf{b}) Based on DMFT, the disordered lattice is mapped onto a set of disorder configurations on a single impurity site, whose Green’s function is computed by coupling the impurity to an effective bath via a self-consistently determined hybridization function $\Delta(t,t')$.
(\textbf{c}) Within the CPA framework, we average over all disorder configurations to obtain the averaged Green’s function $G_{\mathrm{ave}}(t,t')$, which constitutes the DMFT+CPA solution. The self-consistency loop between panels (\textbf{b}) and (\textbf{c}) is iterated until $G_{\mathrm{ave}}(t,t')$ converges.}
\label{fig2.1.1}
\end{figure}

\noindent To study the nonequilibrium dynamics of the Anderson--Hubbard model, we employ our recently introduced method that combines the nonequilibrium extensions of the dynamical mean-field theory (DMFT)~\cite{NeqDMFT1, NeqDMFT2, NeqDMFT3} and that of the coherent potential approximation (CPA)~\cite{NeqCPA1, NeqCPA2} into a solution that we have called DMFT+CPA \cite{NeqDMFT_CPABinaryDisorder, NeqDMFT_CPA2}. The method is schematically summarized in Fig.~\ref{fig2.1.1}. Both DMFT and CPA self-consistently reduce the many-body lattice problem to an effective single-impurity model that can be solved numerically. DMFT maps the entire lattice model with randomly distributed disorder potentials onto a set of disorder configurations defined on a single impurity site; we need to compute the impurity Green's function $G_{V_i}(t,t')$ for each $V_i$ disorder configuration by solving the corresponding impurity problem. We then average over all configurations within the CPA framework to obtain $G_{\text{ave}}(t,t')$. This averaged Green’s function $G_{\text{ave}}(t,t')$, known as the DMFT+CPA solution, incorporates the effects of both disorder and interactions, thus it characterizes the whole system. Note that the self-consistency loop between Fig.~\ref{fig2.1.1}(b) and Fig.~\ref{fig2.1.1}(c) is performed by updating the hybridization function $\Delta(t,t')$ until the convergence of $G_{\text{ave}}(t,t')$ is reached.

\begin{figure}[htbp]
\centering
\includegraphics[width=0.9\columnwidth]{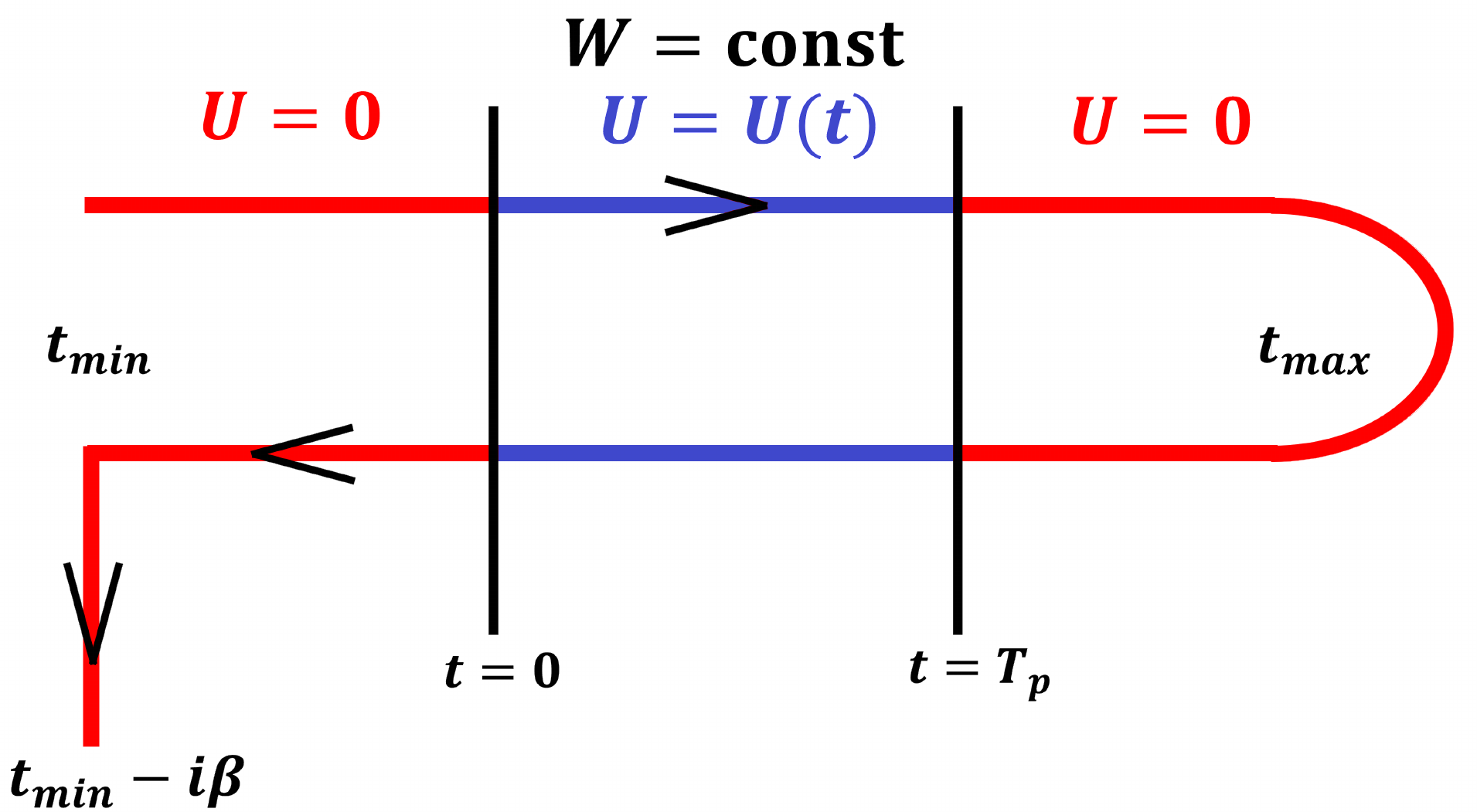}
\caption{The Kadanoff--Baym--Keldysh contour: the system is evolved from an initial time $t_{\min}$ forward to $t_{\max}$, back to $t_{\min}$, and then downward along the imaginary time axis to $t_{\min} - i\beta$. The interaction pulse $U(t)$ start at $t=0$ and end at $t=T_p$, where $T_p$ denotes the pulse duration. The disorder strength $W$ is kept constant throughout the evolution. The real and imaginary-time segments of the contour are discretized using step sizes $\Delta t$ and $\Delta \tau$, respectively.}
\label{fig2.1.2}
\end{figure}

The formalism is constructed on the Kadanoff--Baym--Keldysh contour~\cite{keldysh1, keldysh2, keldysh3, keldysh4, kamenev2011, stefanucci2025, rammer2007}, which, due to the absence of Gell-Mann-Low theorem for the nonequilibrium problem, evolves the system from an initial time $t_{\min}$ forward to $t_{\max}$, back to $t_{\min}$, and then downward along the imaginary time axis to $t_{\min} - i\beta$. In this work, the interaction strength $U(t)$ is explicitly turned on and then turned off according to the chosen pulse protocol, as illustrated schematically in Fig.~\ref{fig2.1.2}.

The central object of the formalism is the contour-ordered Green’s function
\begin{equation}
G_{ij,\sigma}(t,t') = -i \langle T_c\, c_{i\sigma}(t) c_{j\sigma}^{\dagger}(t') \rangle,
\end{equation}
where $T_c$ orders operators along the contour. From this quantity, other real-time Green’s functions can be extracted, including the lesser $G^<(t,t')$, greater $G^>(t,t')$, and retarded $G^R(t,t')$ functions.The contour-ordered Green’s function defined on the Kadanoff–Baym–Keldysh contour provides a unified description of equilibrium and nonequilibrium dynamics. Depending on the positions of the time arguments on the contour branches, different real-time components can be obtained. In particular, the lesser and greater Green’s functions describe particle correlations and occupation properties, while the retarded Green’s function characterizes the causal response of the system. These components follow from projecting the contour-ordered Green’s function onto the corresponding segments of the contour and are related through standard analytic relations within the nonequilibrium Green’s function formalism. Detailed description of this framework can be found in ~\cite{kamenev2011, stefanucci2025, rammer2007}.
Within the DMFT+CPA framework, the lattice problem is mapped onto a single impurity problem embedded in a self-consistently determined medium characterized by the hybridization function $\Delta(t,t')$ \cite{NeqDMFT_CPA2}. For a given disorder configuration, the noninteracting impurity Green’s function is
\begin{equation}
\mathcal{G}_{V_i}(t,t') = \left[(i\partial_t + \mu - V_i)\delta_c - \Delta \right]^{-1}(t,t'),
\end{equation}
and the interacting impurity Green’s function is obtained from the Dyson equation,
\begin{equation}
\label{eq:Dyson}
G_{V_i}(t,t') = \left[ (\mathcal{G}_{V_i})^{-1} - \Sigma_{V_i} \right]^{-1}(t,t'),
\end{equation}
where $\Sigma_{V_i}(t,t')$ is the self-energy for the impurity with on-site disorder $V_i$.  
We employ second-order perturbation theory as the impurity solver.Resulting in the self-energy:
\begin{equation}
\Sigma_{V_i}(t,t') = - U(t) U(t')\, G_{V_i}(t,t')^2\, G_{V_i}(t',t).
\end{equation}
The notation $\langle \cdots \rangle_{\{V\}}$ indicates an average over all possible disorder configurations. Following Eq.~\ref{eq:Dyson}, the disorder-averaged Green’s function is obtained by
\begin{equation}
G_{\text{ave}}(t,t') = \langle G_{V_i}(t,t') \rangle_{\{V\}}.
\end{equation}
On the infinite-dimensional Bethe lattice, the hybridization function is updated via
\begin{equation}
\Delta(t,t') = t^{*2}\, G_{\text{ave}}(t,t').
\label{eq:hybridization}
\end{equation}
This defines the self-consistency loop: starting from an initial guess of $\Delta(t,t')$, we compute $G^0_{V_i}$, evaluate $\Sigma_{V_i}$ and $G_{V_i}$ for each disorder configuration, average to obtain $G_{\text{ave}}$, update $\Delta$, and iterate until convergence of $G_{\text{ave}}$ is reached within a prescribed tolerance. 

In the present work, the lattice is taken to be an infinite-dimensional Bethe lattice, which leads to the compact form of the hybridization  (\ref{eq:hybridization}), and simplifies the DMFT self-consistency loop described above. While the Bethe lattice does not correspond to a specific crystal structure, it is widely used in DMFT studies because it captures the essential feature that the electronic self-energy remains local in the infinite dimensional lattice. % As a result, the method captures local dynamical correlations exactly, while nonlocal spatial correlations are neglected. 
Although quantitative details may depend on the lattice geometry and dimensionality, many qualitative features of strongly correlated systems obtained within DMFT are known to remain robust when comparing different lattice structures, especially for lattices with high coordination numbers. Therefore, the Bethe lattice provides a convenient framework for studying the interplay of interactions, disorder, and nonequilibrium dynamics.

The nonequilibrium DMFT+CPA framework establishes an effective medium that simultaneously captures the effects of both electron--electron interaction and disorder in the nonequilibrium regime. 
The applicability of the DMFT+CPA framework relies on the locality of the self-energy and on the effective-medium treatment of disorder. Dynamical mean-field theory becomes exact in the limit of infinite lattice coordination number, where the self-energy is purely local and nonlocal spatial correlations are neglected while local dynamical correlations are fully retained. Within this framework, disorder is incorporated through the coherent potential approximation by averaging over impurity problems corresponding to different local disorder potentials $V_i$, yielding a disorder-averaged Green’s function that defines the effective medium self-consistently. This procedure is know to appropriately capture the effect of weak to moderate disorder strengths. In the present work, the Anderson–Hubbard model is considered on an infinite-dimensional Bethe lattice, for which the DMFT self-consistency condition is valid. Therefore, the combined DMFT+CPA approach provides a consistent description of the nonequilibrium dynamics of interacting electrons in the presence of site-energy disorder.

The Green's function on the Kadanoff--Baym--Keldysh contour are discretized into $(2N_t + N_\tau)$ time points, where $N_t$ denotes the number of steps along each branch of the real-time contour and $N_\tau$ corresponds to the imaginary-time branch. The step sizes are given by $\Delta t = (t_{\text{max}} - t_{\text{min}})/N_t$ for real time and $\Delta\tau = \beta/N_\tau$ for imaginary time. In this work, we use $t_{\text{min}} = -5$, $t_{\text{max}} = 20$, and an initial inverse temperature $\beta_{\text{initial}} = 15$, with a typical choice of $N_\tau = 100$. Contour-ordered quantities such as $G(t,t')$ are represented as square complex matrices $G_{ij}$ of dimension $(2N_t + N_\tau)\times(2N_t + N_\tau)$, where convolutions become matrix multiplications and continuous inverses are replaced by discrete matrix inverses. To analyze the nonequilibrium dynamics, we employ Wigner coordinates $(T_{\text{ave}}, t_{\text{rel}})$, where $T_{\text{ave}}$ represents the average (or effective) time of the system, and Fourier transforms with respect to $t_{\text{rel}}$ provide access to frequency-domain information. Observables such as the distribution function and energy are computed on three different real-time grids, with typical values of $N_t \sim 1000$, and are subsequently extrapolated to the continuum limit $\Delta t \rightarrow 0$ using quadratic Lagrange interpolation.

%%%%energy part%%%%%%%%

% To calculate the energy, we need to take the expectation value of the Hamiltonian. For the kinetic energy, it is more convenient to translate from lattice space to k-space:
% \begin{equation}
% - t \sum_{{\langle ij \rangle},\, \sigma}
% \left\langle
%     c_{i\sigma}^\dagger c_{j\sigma}
%     +
%     c_{j\sigma}^\dagger c_{i\sigma}
% \right\rangle
% \end{equation}
% =
% we need to translate it into k-space, where it become:
% 2
% in this form, now we can just use the greens function to calculate the kinetic energy.
% For the potential energy

%%%%energy part%%%%%%%%

\subsection{Effective temperature}

\noindent For the calculation of the effective temperature, the method introduced in \cite{NeqDMFT_CPABinaryDisorder} is not suitable for the present system. That approach relies on using the equilibrium energy-versus-temperature data to infer the effective temperature of a nonequilibrium system once it has reached its long-time state. However, in the present case, the system is altered by the interaction pulse and is not allowed to equilibrate to through scattering processes due to the interaction. As a result, the potential energy remains unchanged while the kinetic and total energies vary. Because of this, the final state cannot be matched with an equilibrium state simply by equating the total energy to the equilibrium energy--temperature data curve. Therefore, we instead use the fluctuation--dissipation theorem to analyze the final state after the interaction pulse and deduce an effective temperature of the long-time state of the system. 

\begin{figure}[htbp]
\centering
\includegraphics[width=8cm]{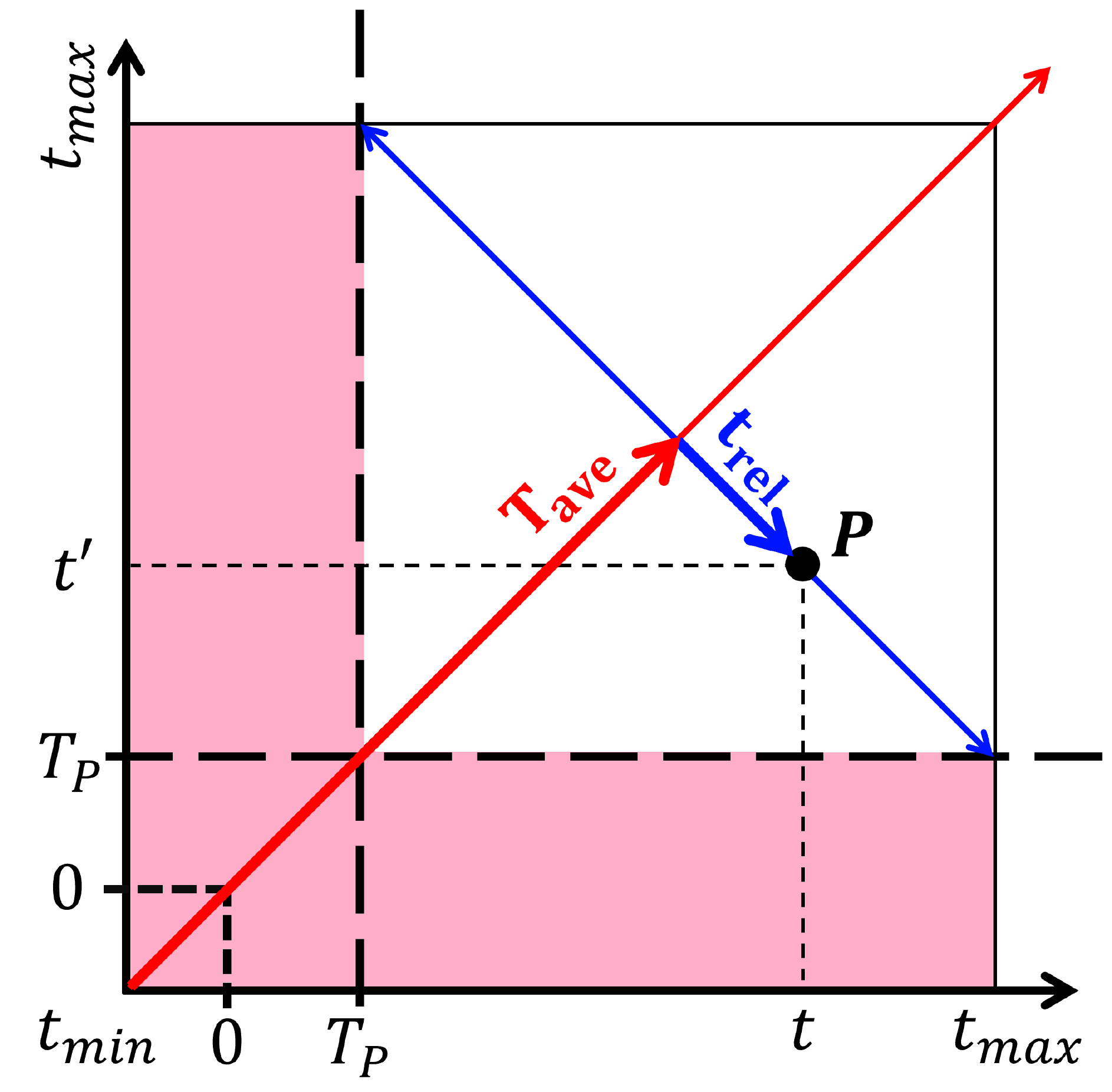}
\caption{ Illustration of the relation between the contour-time coordinates \((t, t')\) and the Wigner coordinates \((T_{\mathrm{ave}}, t_{\mathrm{rel}})\) for a point \(P\) in the two-time space, where \(T_{p}\) denotes the end of the pulse. The pink region indicates the portion of the data that needs to be truncated when evaluating physical quantities in frequency domain after the pulse.
}
\label{fig2.3.1}
\end{figure}

First, because we need to analyze the Green's function in the frequency domain, the time coordinates are transformed from \((t, t')\) to the Wigner coordinates \((T_{\mathrm{ave}}, t_{\mathrm{rel}})\). 
This transformation, illustrated schematically in Fig.~\ref{fig2.3.1}, introduces \(T_{\mathrm{ave}}\), typically interpreted as the effective time of the system, and \(t_{\mathrm{rel}}\), the variable with respect to which Fourier transforms are performed to obtain frequency--space quantities. A Fourier transform of the retarded Green’s function is followed by
\begin{equation}
G^{R}(T_{\mathrm{ave}},\omega)
  = \int dt_{\mathrm{rel}}\, e^{i\omega t_{\mathrm{rel}}}
    G^{R}(T_{\mathrm{ave}}, t_{\mathrm{rel}}),
\end{equation}

To obtain the effective temperature, or its inverse $\beta_{\text{final}}$, after the pulse,  
we apply the nonequilibrium formalism described above to the system in time. The density of states is obtained from the nonequilibrium retarded Green’s function \(G^{R}(T_{\mathrm{ave}},\omega)\):
\begin{equation}
\rho(\omega) = -\frac{1}{\pi}\,\Im\!\left[G^{R}(T_{\mathrm{ave}},\omega)\right].
\end{equation}

In an equilibrium system, the result of this operation is, in principle, independent of $T_{\mathrm{ave}}$, which is not the case for the nonequilibrium system. The value of \(T_{\mathrm{ave}}\) must be carefully chosen such that the available range of \(t_{\mathrm{rel}}\) provides the most reliable numerical evaluation for the Fourier transform (typically taken near the midpoint of the average-time axis). in addition, because the data prior to the end of the pulse are contaminated due to including times $t$ and $t'$  where one is in the pulse and the other after the pulse, the pink region in Fig.~\ref{fig2.3.1} must be truncated. 
To summarize, we choose \(T_{\mathrm{ave}}\) at the midpoint of the remaining average-time interval after the truncation, which is the white area in Fig.~\ref{fig2.3.1}. Based on the fluctuation--dissipation theorem, a thermalized system satisfies
\begin{equation}
G^{<}(\omega) = -2i\, F(\omega)\, \Im G^{R}(\omega),
\end{equation}
where \(F(\omega)\) denotes the distribution function. The effective temperature is obtained by fitting the system’s distribution function to a Fermi--Dirac $F(\omega) = \frac{1}{1 + \exp(\beta \omega)}$
with $\beta$ as a free parameter, which is extracted from the fit over a small frequency window around the Fermi energy $-0.1<\omega < 0.1$. Because the truncation limits the accessible range of $t_{\text{rel}}$, an extension of the time range may be necessary. This consideration will be examined in the Results section.

%%%%%%%%%%%%%%%%%%%%%%%%%%%%%%%%%%%%%%%%%%
\section{Results}
\label{sec:Results}

\subsection{Change in energy for different interaction pulse shapes and disorder}

\noindent We study the nonequilibrium dynamics induced by an interaction pulse, during which the interaction strength \(U(t)\) is switched on and subsequently turned off, starting and ending at zero following a given time-dependent profile. We compute kinetic, potential, and total energies as functions of time \cite{Eckstein2010}. 
From the expression of the kinetic energy per lattice site as
\begin{equation}
E_{\mathrm{kin}}(t)
= \frac{1}{N} \sum_{k,\sigma} \varepsilon_k \,
\left\langle c_{k,\sigma}^\dagger(t)\, c_{k,\sigma}(t) \right\rangle,
\end{equation}
where \(N\) denotes the number of lattice sites, \(k\) the momentum vector, and \(\varepsilon_k\) is the band dispersion, it can therefore be re-expressed as
\begin{equation}
E_{\mathrm{kin}}(t)
= 2 \int d\varepsilon \, \rho(\varepsilon)\,
\varepsilon \, G^{<}(t,t),
\end{equation}
where \(\varepsilon\) is the band energy.
The potential energy is obtained from the expression:
\begin{equation}
E_{\mathrm{pot}}(t)
= \bigl[ G_{\mathrm{ave}} * \Sigma_{\mathrm{ave}} \bigr]^{<}(t,t)
+ \frac{U(t)}{4}.
\end{equation}
The total energy can be obtained directly by summing up the kinetic and potential energies.

\begin{figure}[htbp]
\centering
\subfigure{\includegraphics[width=0.49\linewidth]{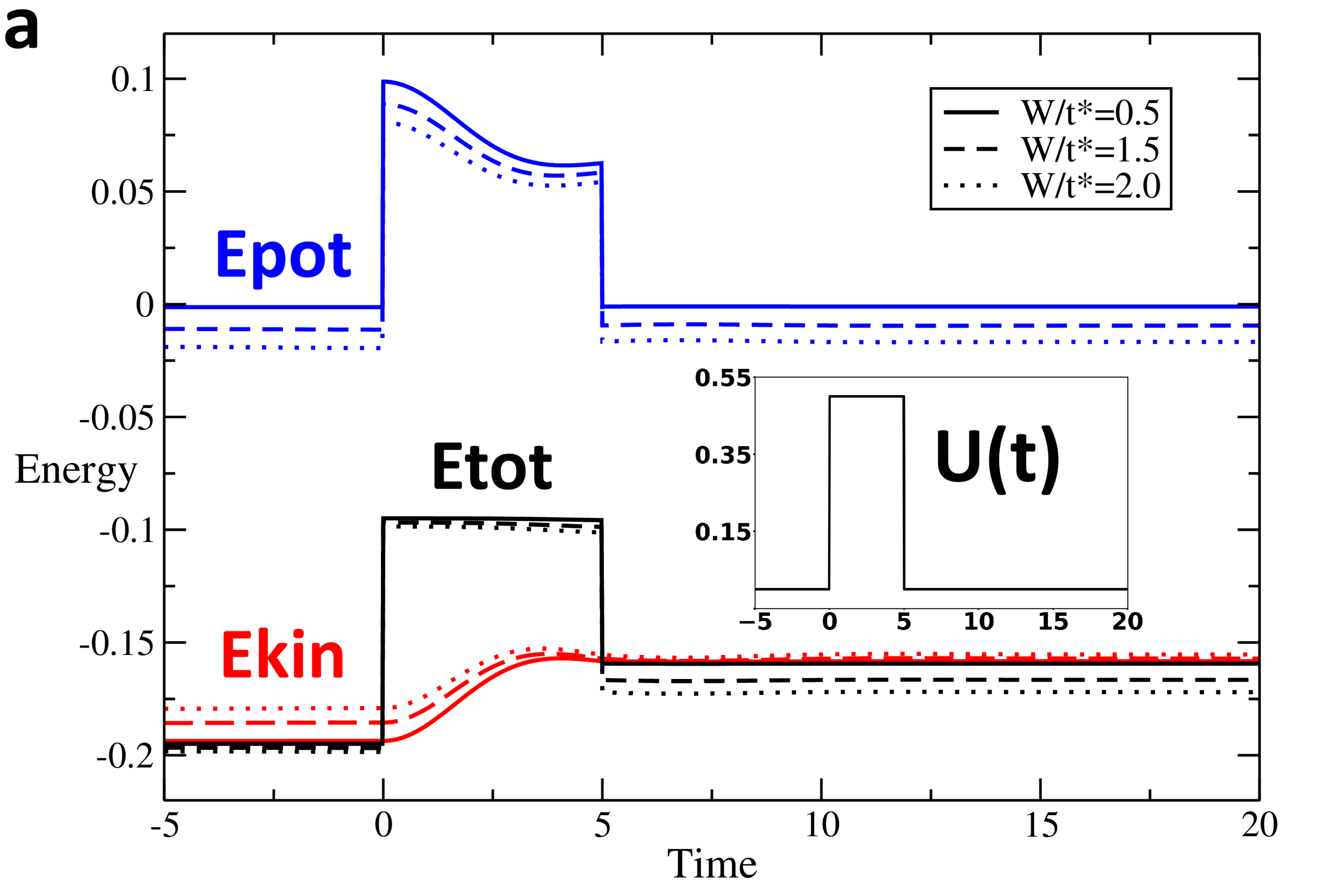}}
\hfill
\subfigure{\includegraphics[width=0.49\linewidth]{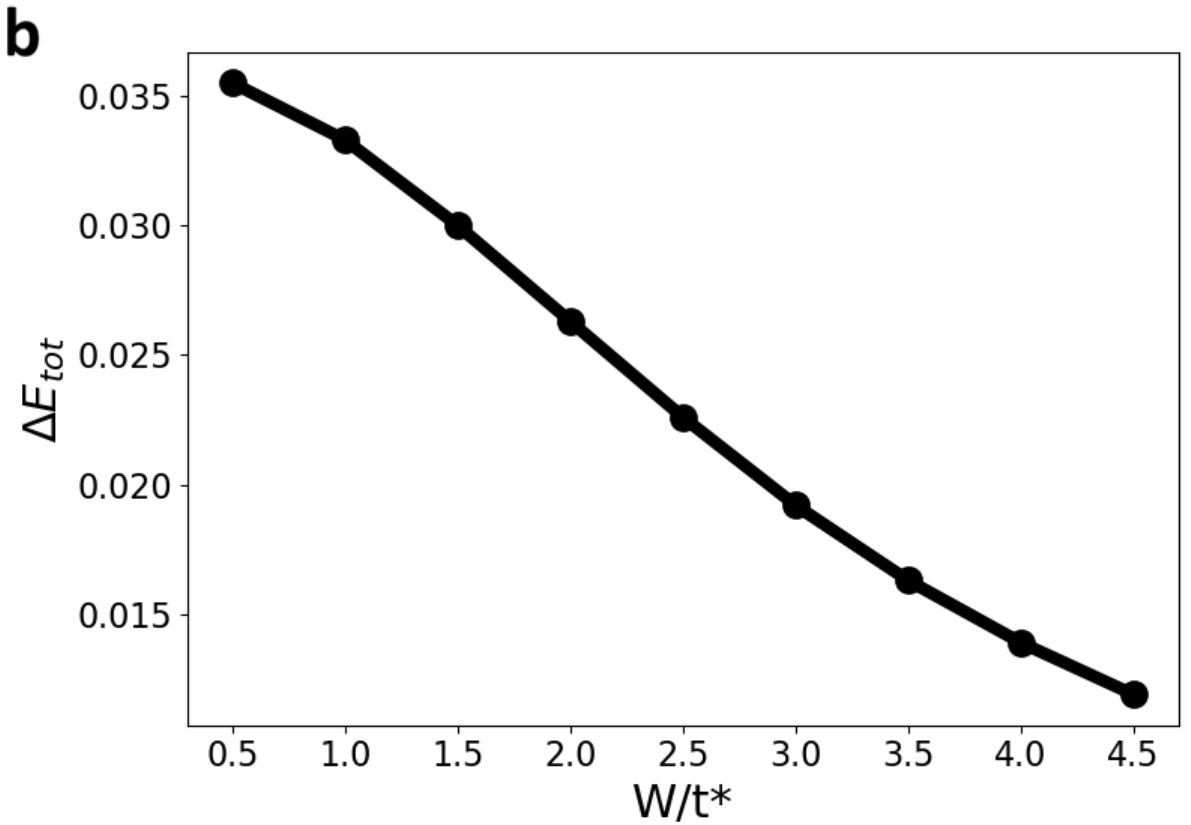}}
\caption{For the rectangular pulse interaction ($U_{\text{max}}/t^* = 2$, pulse width $T_{\text{p}}$=5), 
the change in total energy $\Delta E_\text{tot}$ systematically decreases as the disorder strength W increases, 
indicating that disorder promotes adiabaticity for this sharp on--off driving: (\textbf{a}) Energy as function of time for rectangular pulse interaction with $U_{\text{max}}/t^* = 2$, and pulse width $T_{\text{p}} = 5$ for $W/t* = \; 0.5, \; 1.5, \;2$. The blue, red, and black lines represent the potential, kinetic, and total energy, respectively. (\textbf{b}) $\Delta E_\text{tot}$ vs W for Rectangular Pulse with $U_{\text{max}}/t* = 2$, and pulse width $T_{\text{p}} = 5$. $\Delta E_\text{tot}$ is defined as $E_{\text{tot}}(t_\text{max})-E_{\text{tot}}(t_\text{min})$}.
\label{fig:fig3.1.1}
\end{figure}

\begin{figure}[htbp]
\centering
\subfigure{\includegraphics[width=0.49\linewidth]{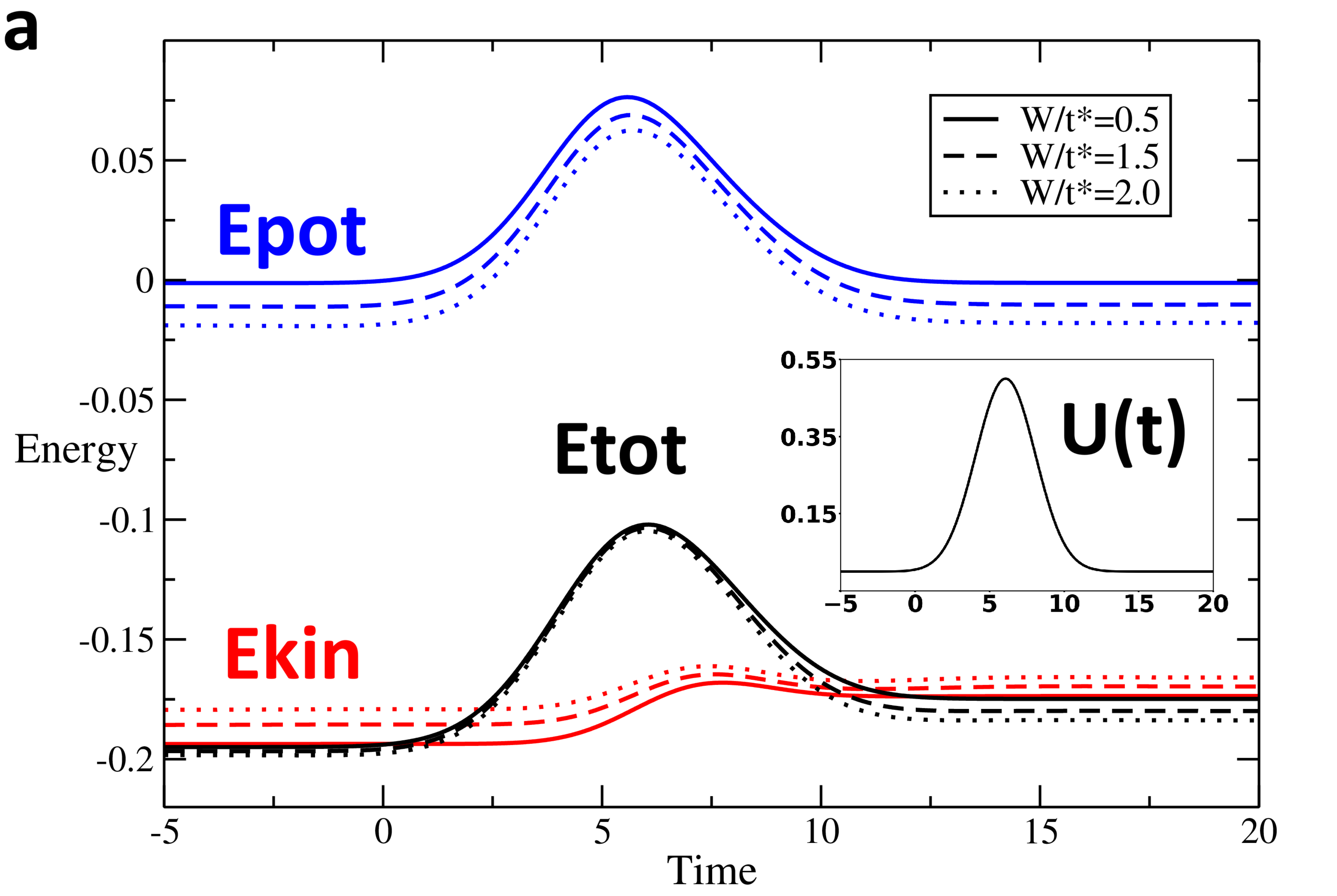}}
\hfill
\subfigure{\includegraphics[width=0.49\linewidth]{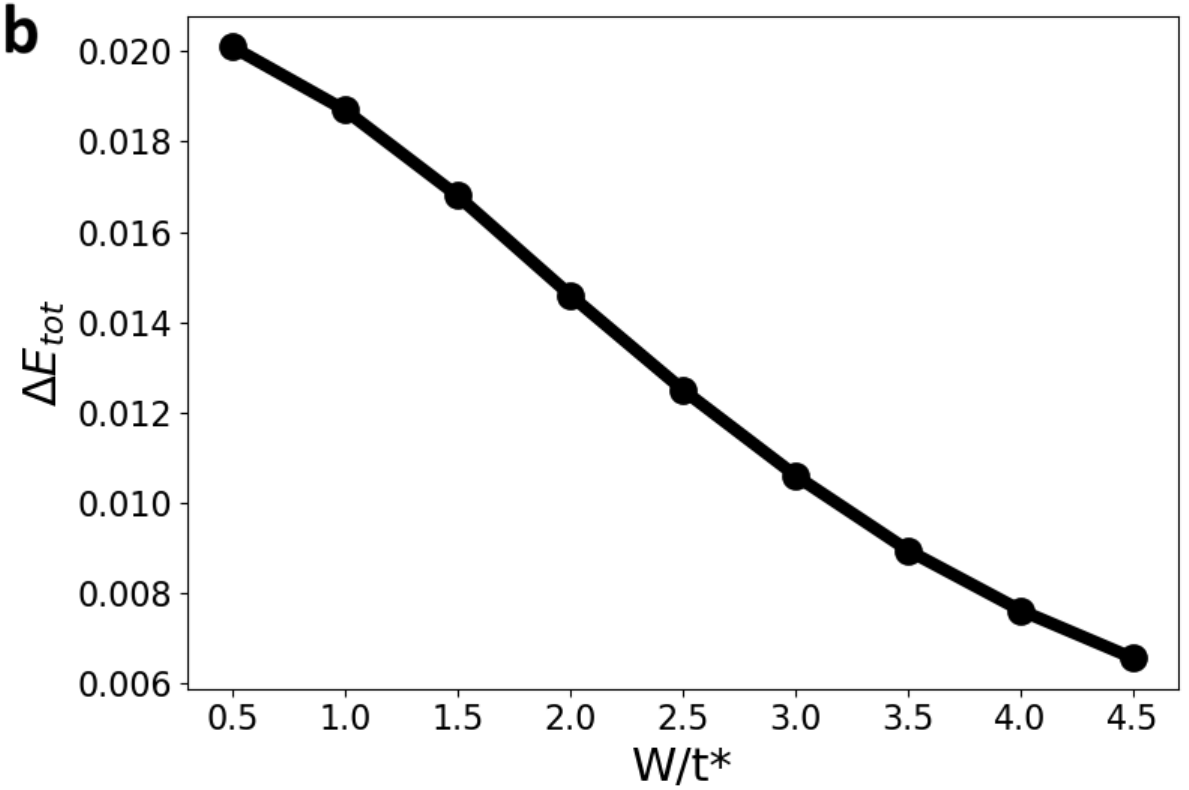}}
\caption{For the Gaussian pulse interaction ($U_{\text{max}}$/t*=2, pulse width $T_{\text{p}}$=12.14), 
$\Delta E_\text{tot}$ again decreases with increasing disorder. 
In this smoothly varying protocol, disorder enhances the adiabatic response in agreement with the rectangular and triangular cases: (\textbf{a}) Energy as function of time for Gaussian pulse interaction with $U_{\text{max}}$/t*=2, and pulse width $T_{\text{p}}$=12.14 for W/t*=0.5, 1.5, 2. (\textbf{b}) $\Delta E_\text{tot}$ vs W for gaussian Pulse with $U_{\text{max}}$/t*=2, and pulse width $T_{\text{p}}$=12.14.}
\label{fig:fig3.1.2}
\end{figure}

\begin{figure}[htbp]
\centering
\subfigure{\includegraphics[width=0.49\linewidth]{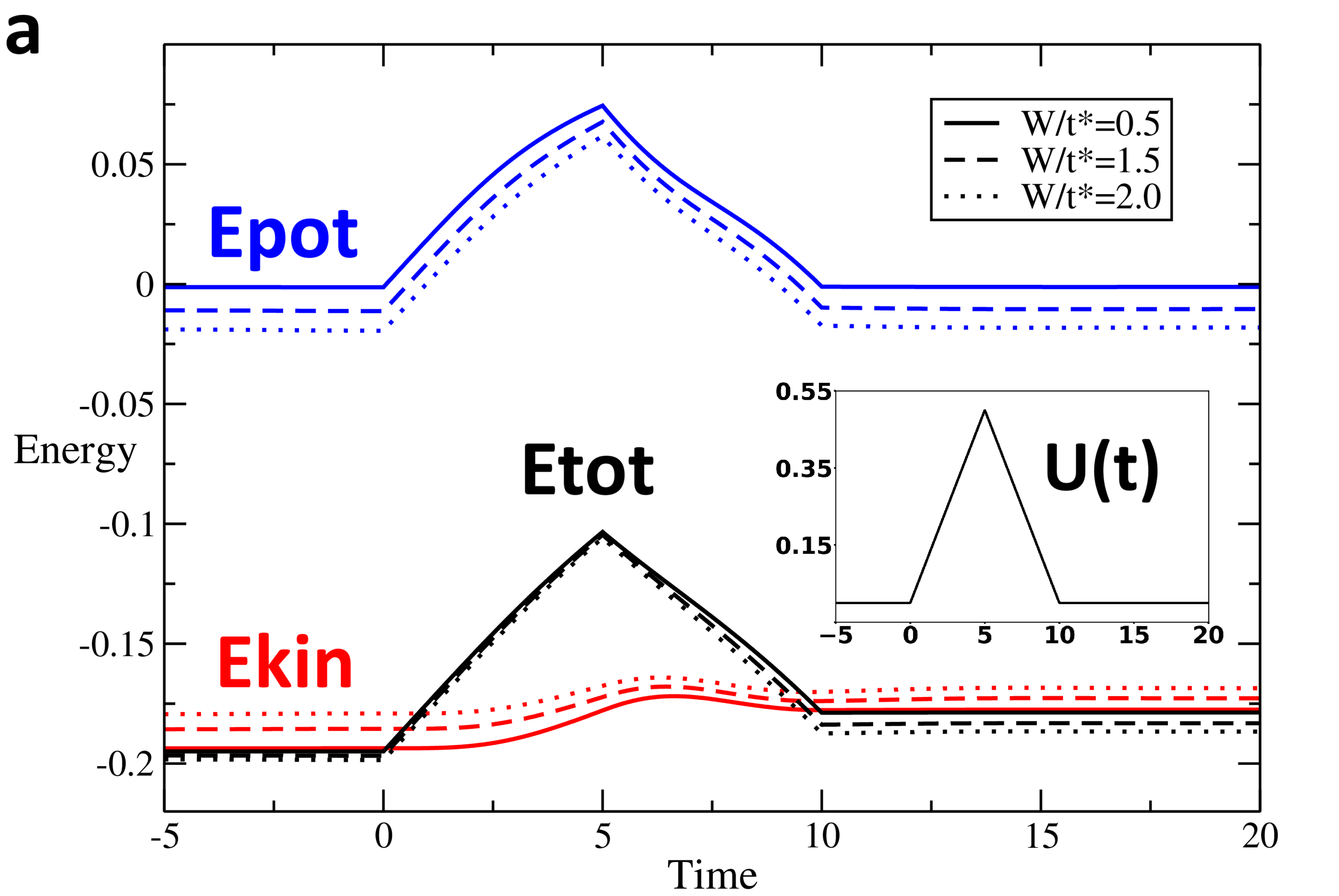}}
\hfill
\subfigure{\includegraphics[width=0.49\linewidth]{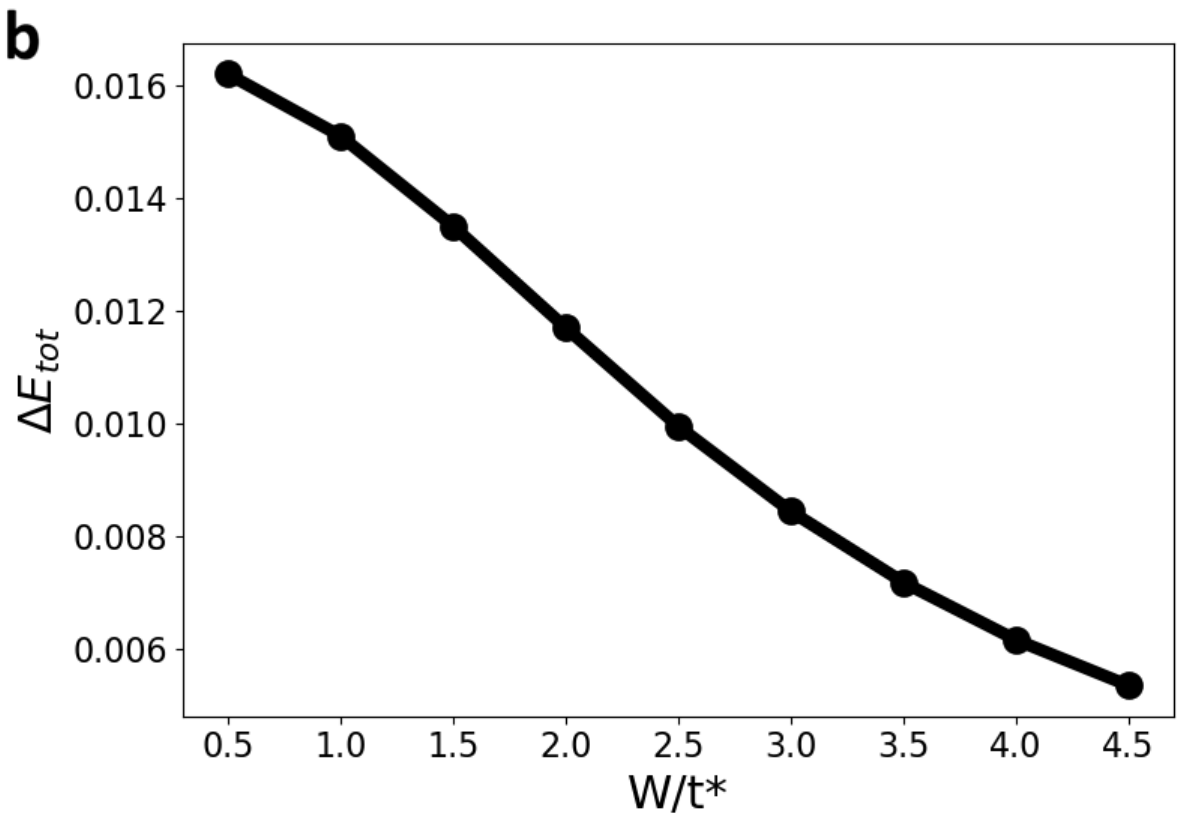}}
\caption{For the triangular pulse interaction ($U_{\text{max}} /t^* = 2$, pulse width $T_{\text{p}} = 10$), 
a similar decrease of $\Delta E_\text{tot}$ with increasing disorder is observed, 
showing that disorder also enhances adiabaticity in a gradual ramping protocol: (\textbf{a}) Energy as function of time for triangular pulse interaction with $U_{\text{max}}/t^* = 2$, and pulse width $T_{\text{p}} = 10$ for $W/t^* = 0.5, \; 1.5, \; 2$. (\textbf{b}) $\Delta E_\text{tot}$ vs W for Triangular Pulse with $U_{\text{max}} /t^* = 2$, and pulse width $T_{\text{p}} = 10$.}
\label{fig:fig3.1.3}
\end{figure}

We track the variation of these different types of energy across the interaction modulation for different profiles and as a function of disorder. Figures (\ref{fig:fig3.1.1})--(\ref{fig:fig3.1.3}) display the time evolution, shown in the (a) panels, of the potential (blue), kinetic (red), and total (black) energies for rectangular, Gaussian, and triangular pulses respectively, for disorder strengths $W = 0.5 t^* $ (solid line), $W = 1.5 t^* $ (dashed line), $W = 2 t^* $ (dotted line). After the interaction returns to zero, the potential energy also vanishes and returns to its initial value. However, the kinetic energy after the pulse is different from its initial value, leading to a similar behavior of the total energy. The (b) panels of the figures show the change in total energy across the interaction pulse as a function of the disorder strength. Across all pulse shapes, we see that the change in total energy is suppressed by an increased disorder strength.

\begin{figure}[htbp]
\centering
\subfigure{\includegraphics[width=0.49\linewidth]{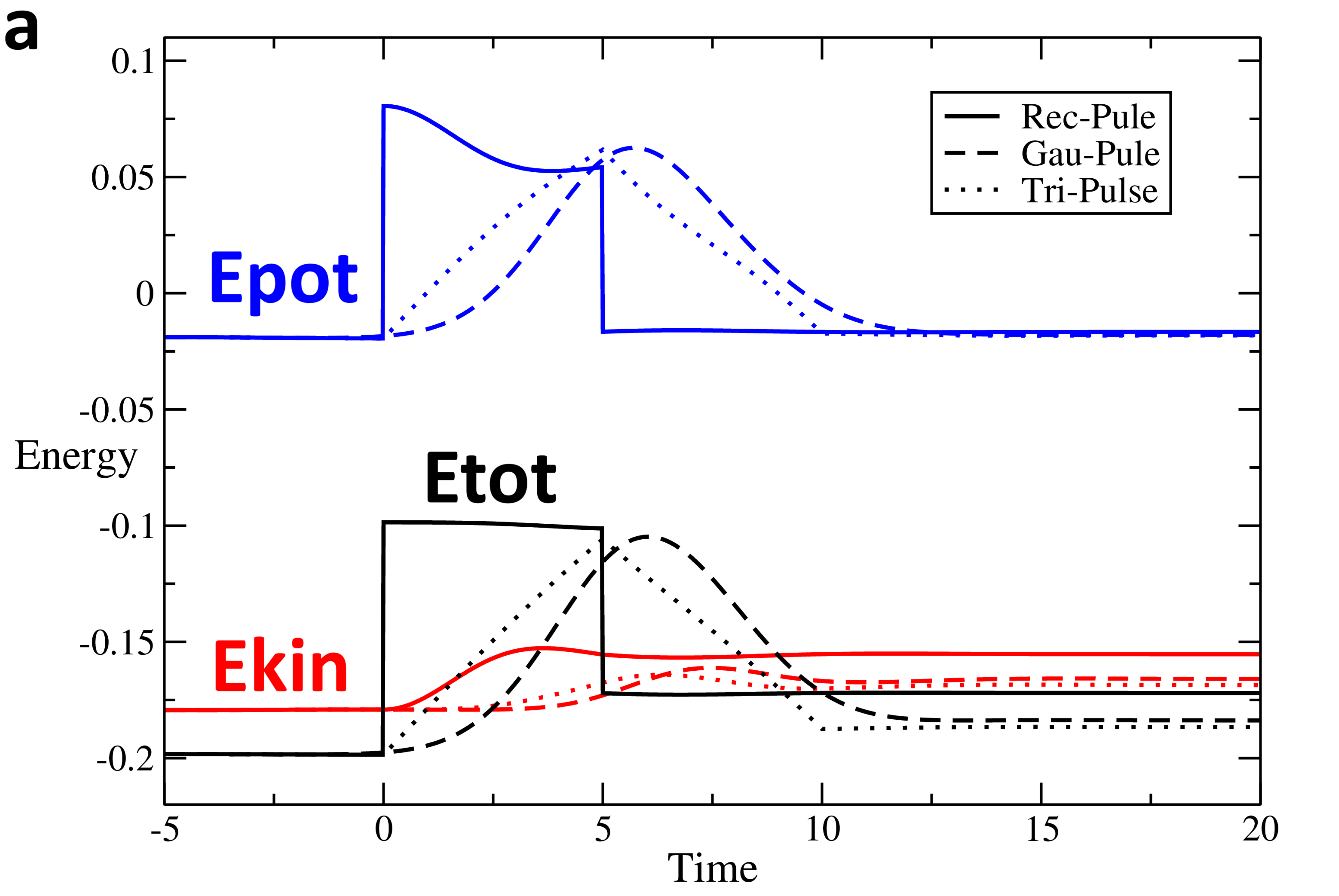}}
\hfill
\subfigure{\includegraphics[width=0.49\linewidth]{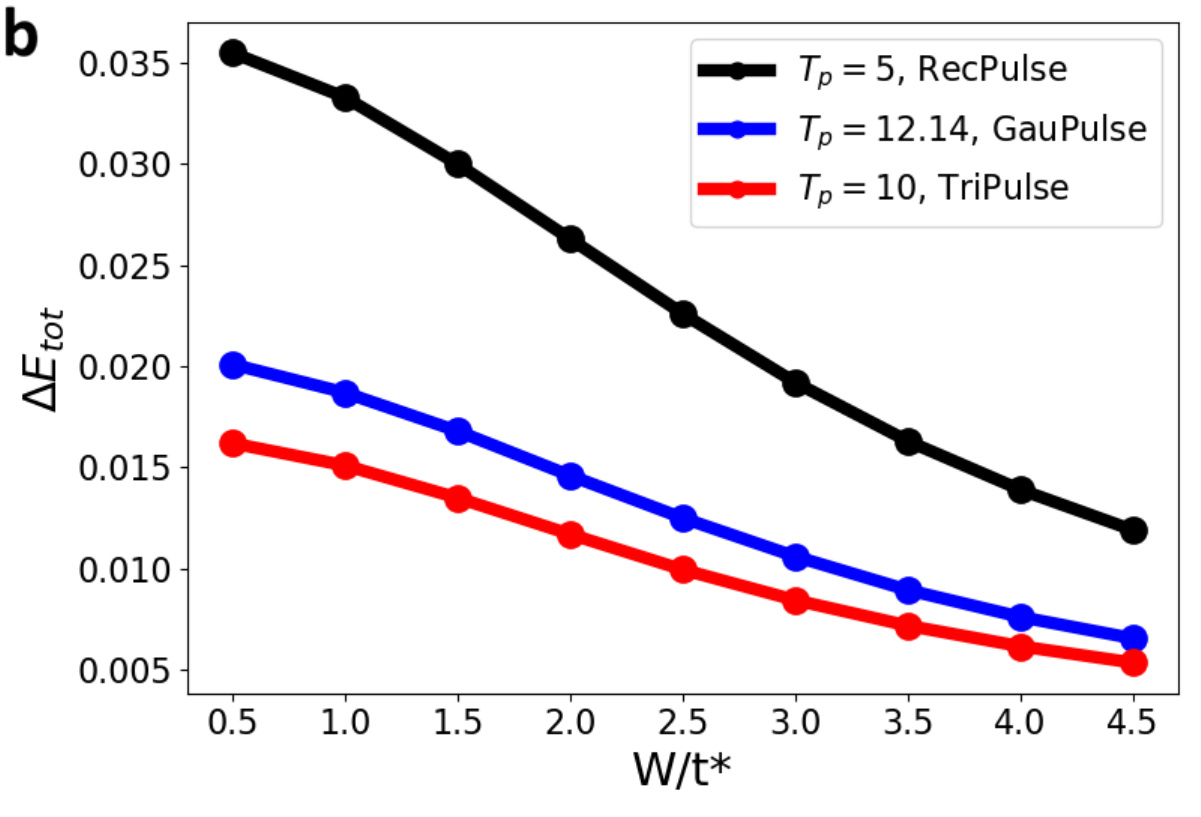}}
\caption{For the rectangular, triangular and gaussian pulses interaction ($U_{\text{max}}$/t*=2, equal area), 
$\Delta E_\text{tot}$ always decreases with increasing disorder strength W. This shows that disorder enhances the adiabaticity of the process regardless of the type of interaction pulse applied: (\textbf{a}) Energy as function of time for rectangular, triangular and gaussian pulse interaction with $U_{\text{max}}$/t*=2 and equal area (different $T_{\text{p}}$) at W/t*=2. (\textbf{b}) $\Delta E_\text{tot}$ vs W/t* for three pulse types with $U_{\text{max}}$/t*=2 and equal area.}
\label{fig:fig3.1.4}
\end{figure}

\begin{figure}[htbp]
\centering
\subfigure{\includegraphics[width=0.49\linewidth]{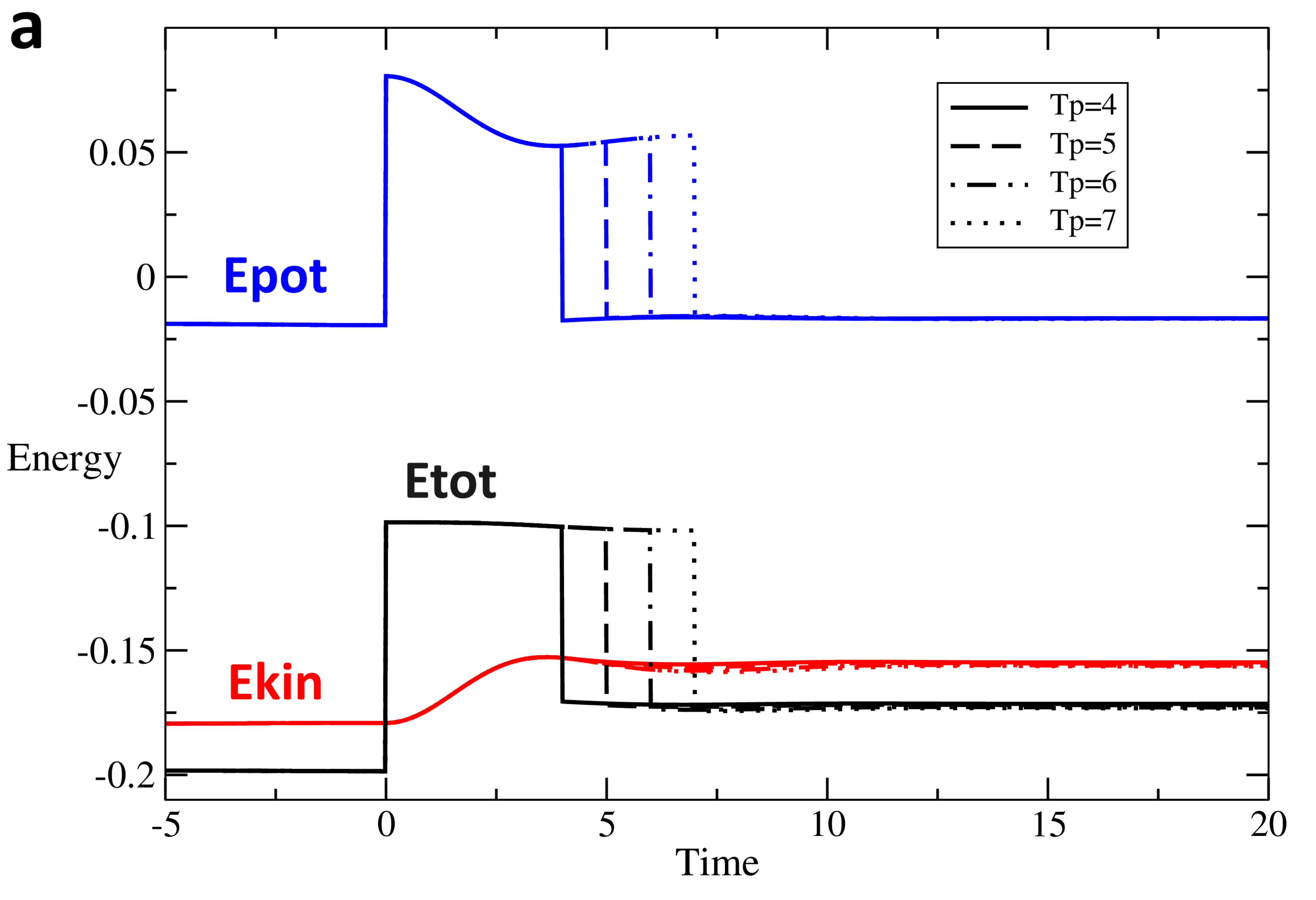}}
\hfill
\subfigure{\includegraphics[width=0.49\linewidth]{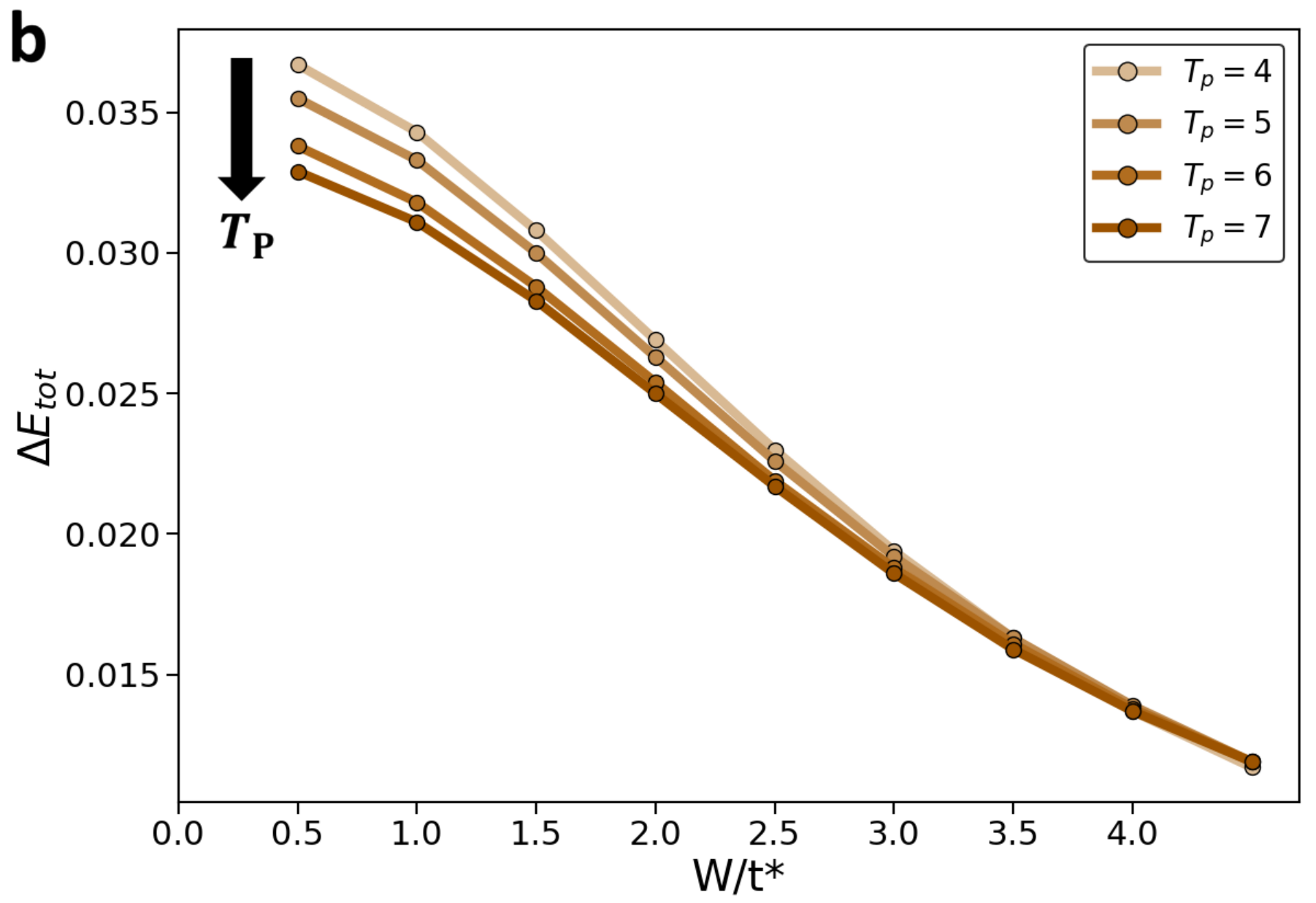}}
\caption{For the rectangular interaction pulse ($U_{\text{max}}$/t*=2, W/t*=2), 
$\Delta E_\text{tot}$ decreases as the pulse period $T_{\text{p}}$ increases, indicating that a longer period promotes adiabaticity in the rectangular pulse interaction: (\textbf{a}) Energy as function of time for rectangular pulse interaction with $U_{\text{max}}$/t*=2 and W/t*=2 for pulse width $T_{\text{p}}$= 4, 5, 6, 7. The blue, red, and black lines represent the potential, kinetic, and total energy, respectively. (\textbf{b}) $\Delta E_\text{tot}$ vs W for rectangular pulse, colors transition from light brown to dark brown corresponding to pulse width $T_{\text{p}}$=4, 5, 6, 7 respectively.}
\label{fig:fig3.2.1}
\end{figure}

Fig.~\ref{fig:fig3.1.4} shows a comparison in the time evolution of the energies for different pulse shapes, with the same area under the curve, for disorder strength $W/t^* = 2$ in panel (a). In panel (b), it presents the change in total energy as a function of disorder strength for the rectangular pulse (black), for the Gaussian pulse (red) and for the triangular pulse (blue) as a function of the disorder strength. We observe that, for all pulse shapes, increased disorder suppresses the change in total energy, thus showing a more adiabatic response. Moreover, the triangular pulse protocol is seen to yield the smallest $\Delta E_\text{tot}$.
%, demonstrating that this protocol leads to the most adiabatic response.

\begin{figure}[htbp]
\centering
\subfigure{\includegraphics[width=0.49\linewidth]{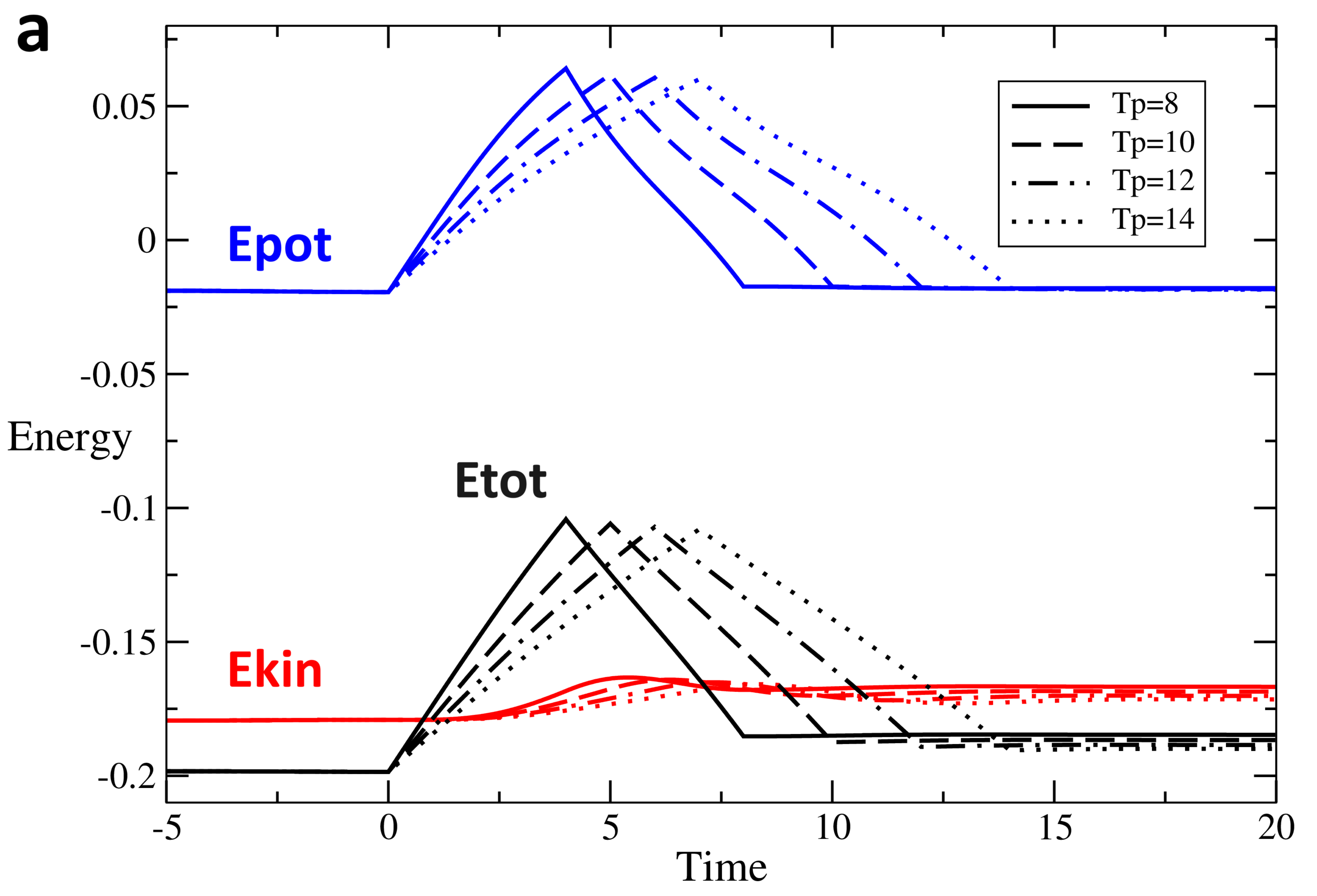}}
\hfill
\subfigure{\includegraphics[width=0.49\linewidth]{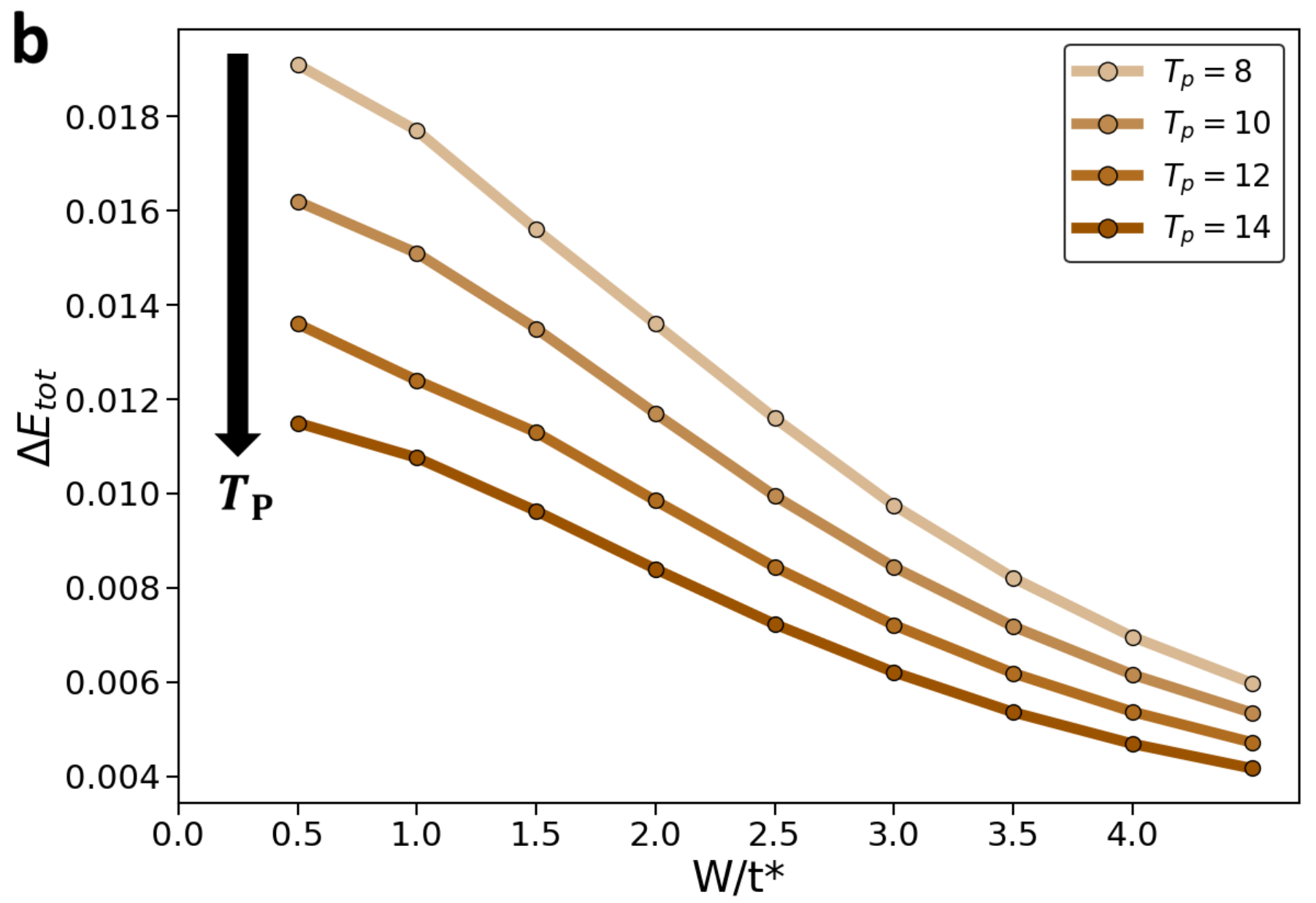}}
\caption{For the Triangular interaction pulse ($U_{\text{max}}$/t*=2, W/t*=2), 
$\Delta E_\text{tot}$ decreases with increasing pulse period $T_{\text{p}}$, showing that longer pulse width enhance adiabaticity for the triangular driving protocol: (\textbf{a}) Energy as a function of time for triangular pulse interaction with $U_{\text{max}}$/t*=2 and W/t*=1.0 for pulse width $T_{\text{p}}$= 8, 10, 12, 14. (\textbf{b}) $\Delta E_\text{tot}$ vs W for triangular pulse, colors transition from light brown to dark brown corresponding to pulse width $T_{\text{p}}$= 8, 10, 12, 14 respectively.}
\label{fig:fig3.2.2}
\end{figure}

\begin{figure}[htbp]
\centering
\subfigure{\includegraphics[width=0.49\linewidth]{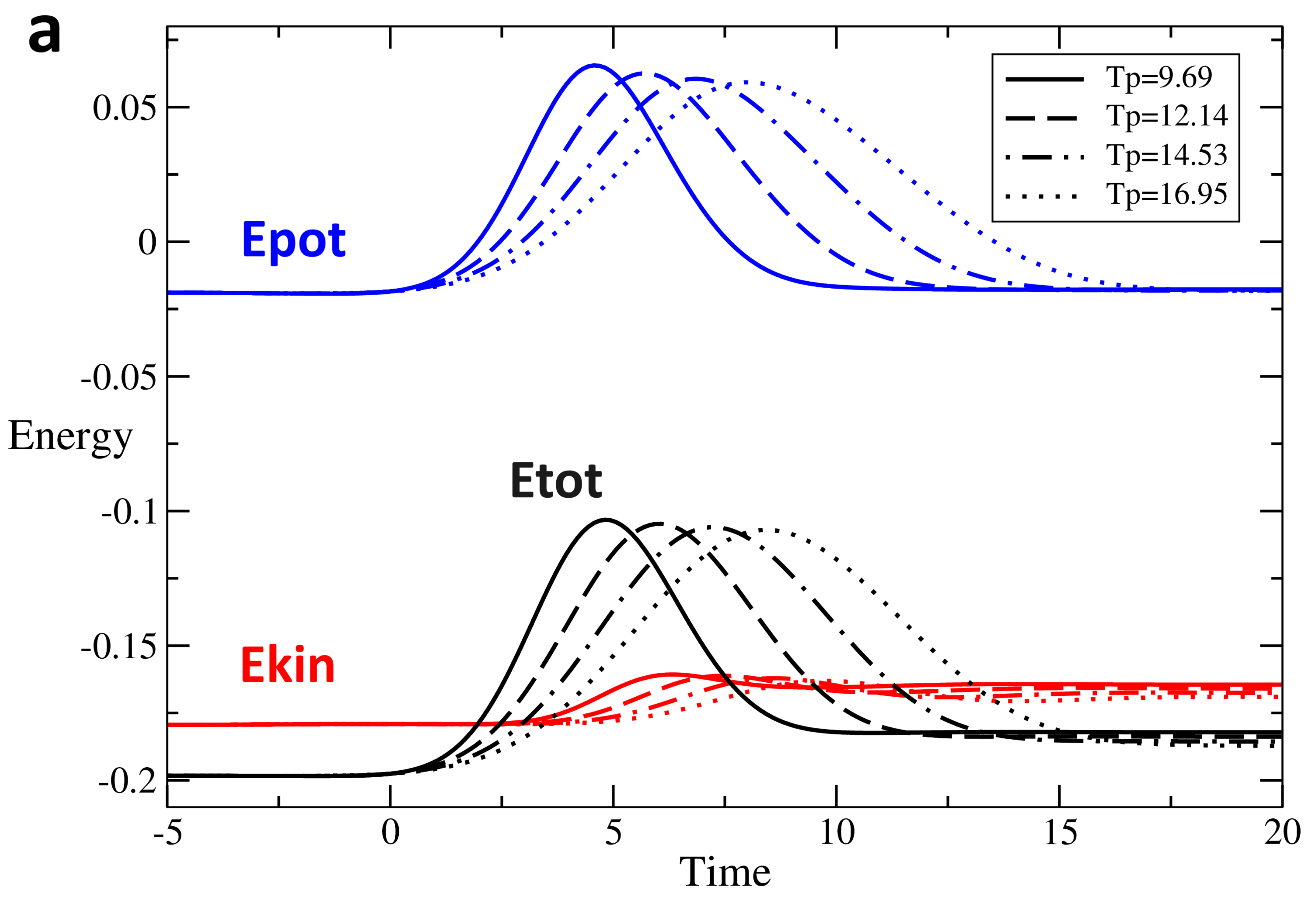}}
\hfill
\subfigure{\includegraphics[width=0.49\linewidth]{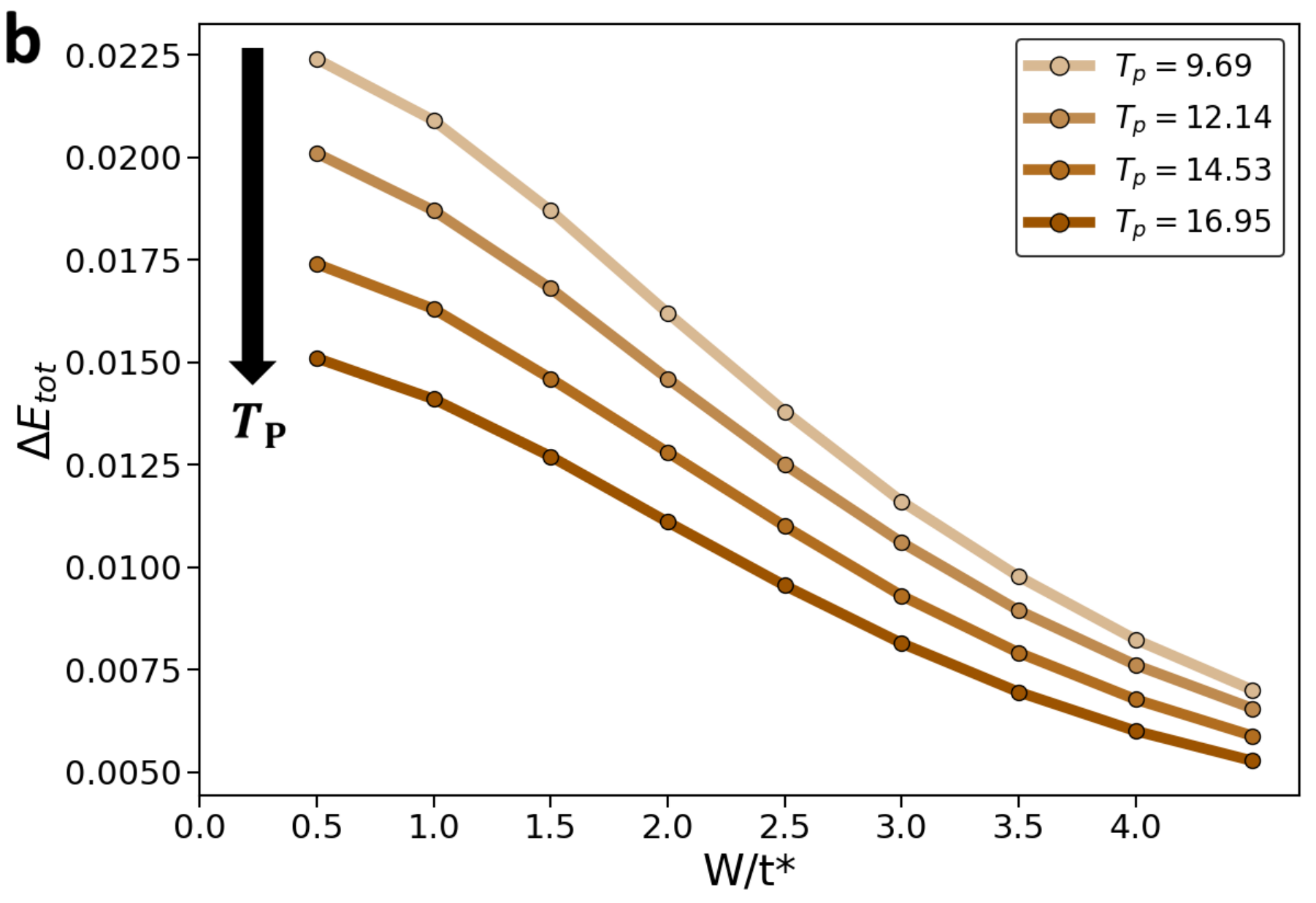}}
\caption{For the gaussian interaction pulse ($U_{\text{max}}$/t*=2, W/t*=2), 
$\Delta E_\text{tot}$ also decreases as the width $T_{\text{p}}$ grows, demonstrating that extended periods improve adiabaticity in the gaussian pulse case: (\textbf{a}) Energy as a function of time for gaussian pulse interaction with $U_{\text{max}}$/t*=2 and W/t*=2 for width $T_{\text{p}}$= 9.69, 12.14, 14.53, 16.95. (\textbf{b}) $\Delta E_\text{tot}$ vs W for gaussian pulse, colors transition from light brown to dark brown corresponding to pulse widths $T_{\text{p}}$= 9.69, 12.14, 14.53, 16.95 respectively.}
\label{fig:fig3.2.3}
\end{figure}

\subsection{Change in energy for different pulse durations and disorder}

\noindent Figures (\ref{fig:fig3.2.1}), (\ref{fig:fig3.2.2}) and (\ref{fig:fig3.2.3}) show in the (a) panels, respectively for the rectangular, the triangular and the Gaussian pulse respectively, the time evolution of the potential energy (blue), kinetic energy (red) and total energy (black) for different pulse durations $T_p$. The (b) panels of the figures show the change in total energy, $\Delta E_{\text{tot}}$, across the interaction pulse as a function of the disorder strength for different pulse durations. While we see that pulse duration has the weakest effect on $\Delta E_{\text{tot}}$ for the rectangular pulse, a broader pulse leads to a smaller $\Delta E_{\text{tot}}$. Most importantly, we observe that increased disorder strength has a stronger effect in producing a  more adiabatic response than the pulse duration, eventually overwhelming the effect of the pulse duration for moderate to strong disorder strength. 

%Across all pulse shapes, we see that the change in total energy is suppressed by an increased disorder strength.
%Besides studying $\Delta E_{\text{tot}}$ as a function of the disorder strength $W$ for different pulse shapes, we also investigate its dependence on $W$ for different pulse periods $T_{\text{p}}$. In general, a longer pulse period $T_{\text{p}}$ corresponds to a gentler driving protocol. As shown in Figs.~\ref{fig3.2.1}, \ref{fig3.2.2}, and \ref{fig3.2.3}, the curves shift downward as $T_{\text{p}}$ increases, indicating that $\Delta E_{\text{tot}}$ decreases with increasing $T_{\text{p}}$. This demonstrates a negative correlation between $\Delta E_{\text{tot}}$ and $T_{\text{p}}$.

The change in the total energy following the interaction modulation reflects the degree of nonadiabaticity across the pulse. When the interaction strength is varied in time, the system absorbs energy as the modulation drives transitions between many-body states. The magnitude of this energy increase depends on the rate of interaction change and this rate varies with the pulse shape.    % how the interaction is distributed in time. The pulse shapes differ in the rate of increase of the interaction strength. In particular, different pulse shapes distribute the interaction strength differently during the modulation. 
The change of the interaction is instantaneously tied to the potential energy thus the change in potential energy immediately vanishes when the interaction is returned to zero. However, the kinetic energy is dynamically adjusted and would eventually settled into a steady state value if the system were allowed to equilibrate over a long time with a finite interaction strength.
Disorder further modifies this behavior by suppressing coherent electron motion and reducing the ability of the system to absorb energy from the interaction modulation. As the disorder strength increases, it limits the generation of nonequilibrium excitations, the kinetic energy change across the pulse is reduced and leads to a smaller change in the total energy. These combined effects explain the trends observed in our numerical results for different pulse protocols, durations, and disorder strengths.
Overall, pulse protocols that keep the interaction near its maximum value for longer intervals tend to produce larger values of $\Delta E_{\mathrm{tot}}$ and stronger nonadiabatic effects. The pulse duration also plays an important role: longer pulses allow the system to adjust more gradually to the time-dependent interaction through a redistribution of the additional potential energy into kinetic energy and therefore approach a more adiabatic evolution, resulting in smaller energy absorption when the interaction is returned to zero.

\begin{figure}[htbp]
\centering
\subfigure{\includegraphics[width=0.49\linewidth]{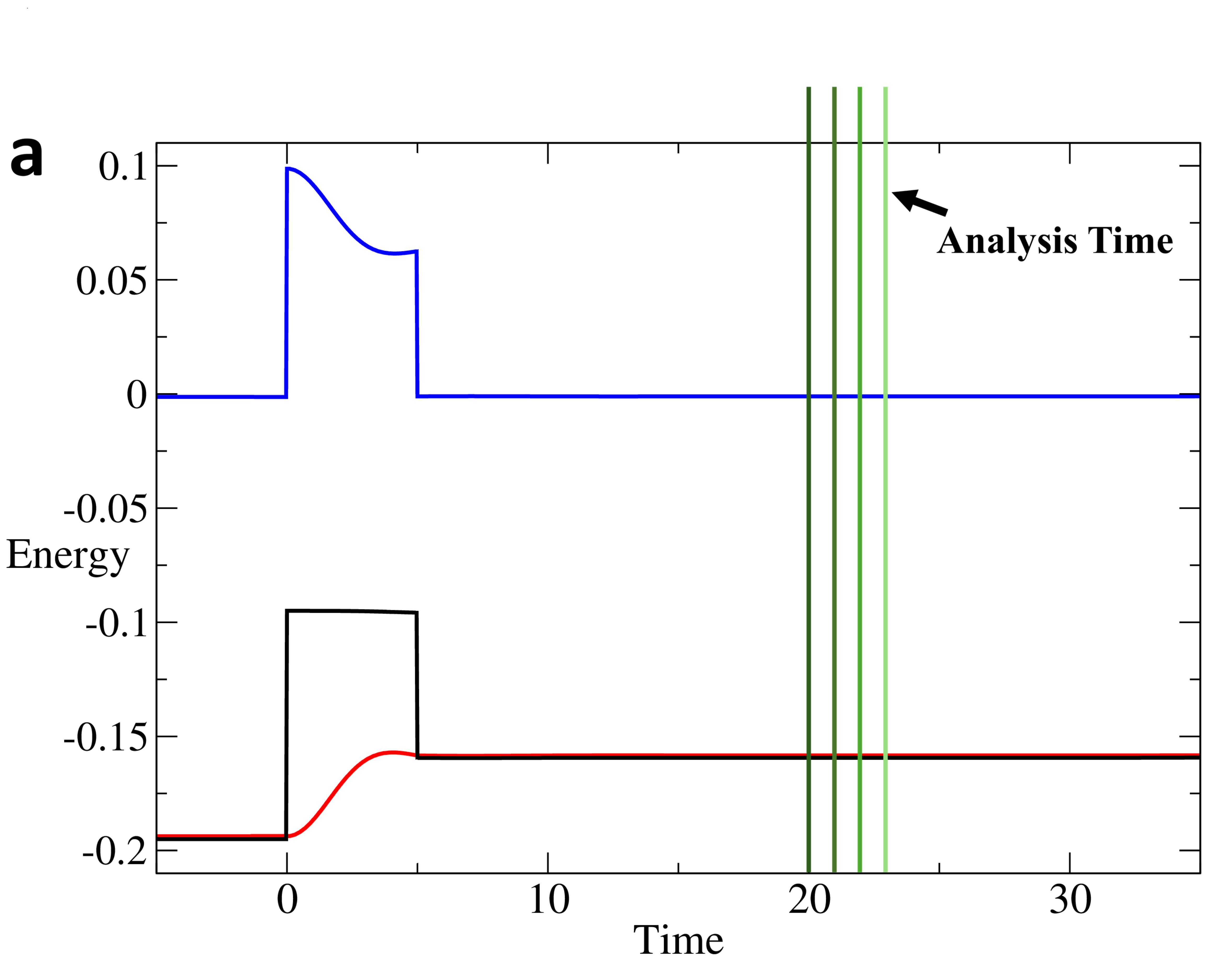}}
\hfill
\subfigure{\includegraphics[width=0.49\linewidth]{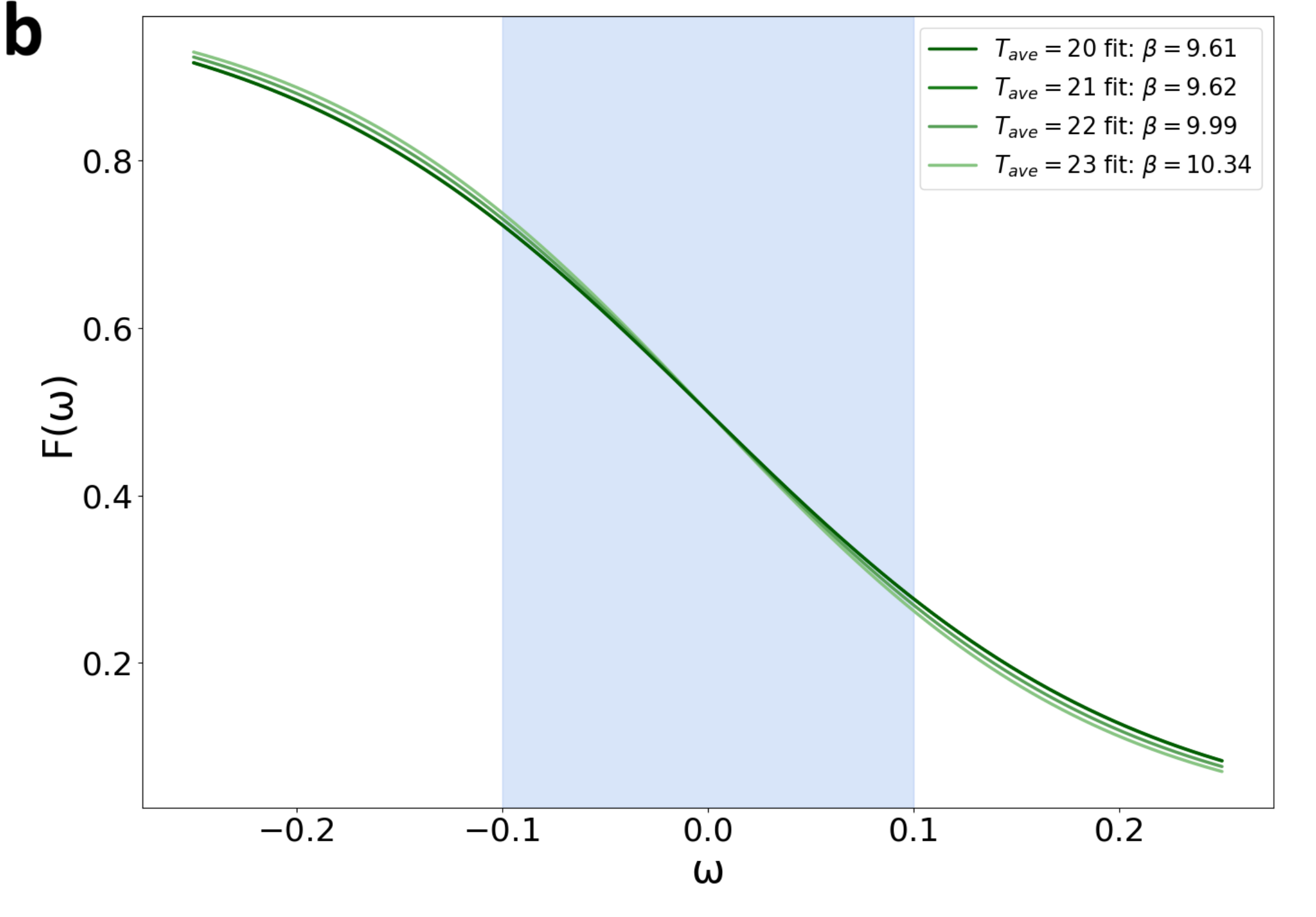}}
\caption{Effective temperature analysis for the rectangular pulse interaction with $U_{\text{max}}$/t*=2, pulse width $T_{\text{p}}=5$, $\beta_{\text{initial}}=15$ and $W/t^* = 0.5$. the time scale range is from -5 to 35 (\textbf{a}) The green lines represent the analysis time $T_{\text{ave}}$. (\textbf{b}) Post-relaxation \(F(\omega)\) for $T_{\text{ave}}=20$ (midpoint of the residual interval) and slightly later $T_{\text{ave}}$.}
\label{fig:fig3.3.1}
\end{figure}

\subsection{Effective final temperature vs. U for Different Pulse shape and W}

\noindent Finally, we investigate the evolution of the effective temperature in the system under the time modulation of the interaction. As previously noted, the numerical analysis of the effective temperature after the pulse requires a careful choice of the average time $T_{\text{ave}}$ and a sufficient range of relative time $t_{rel}$ to perform a reliable Fourier transform for the application of the fluctuation dissipation theorem. Therefore, care is taken to appropriately choose the parameters for the extraction of the effective temperature of the system.

%We start by performing this check in Figs.~\ref{fig:fig3.3.1} and \ref{fig:fig3.3.2}.

\begin{figure}[htbp]
\centering
\includegraphics[width=\columnwidth]{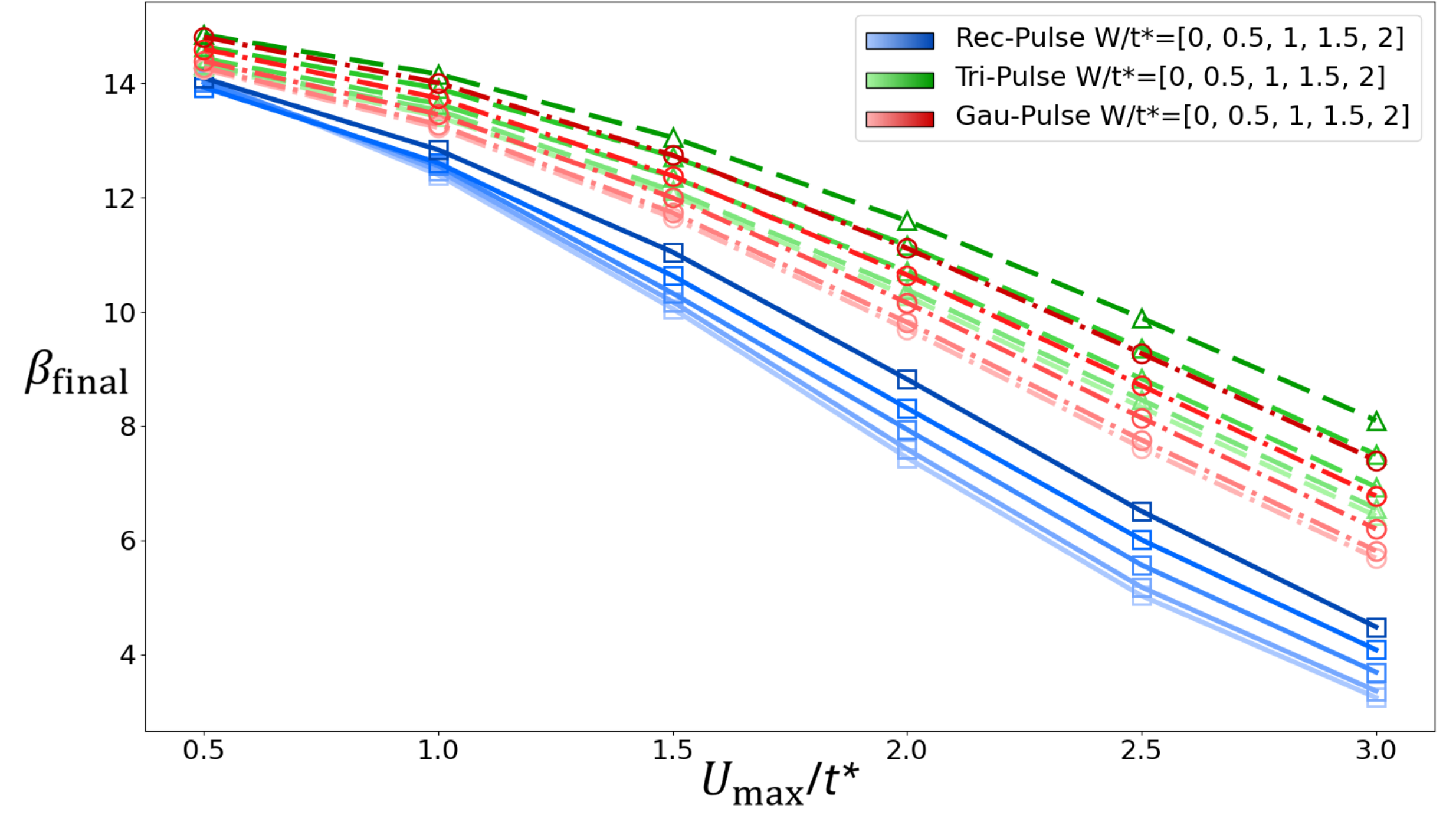}
\caption{ Inverse of the final effective temperature $\beta_{\text{final}}$ as a function of interaction strength $U/t^*$ is shown for different pulse shapes and disorder strengths $W$; the temperature before the pulse is $\beta_{\text{initial}}=15$. The data for rectangular, triangular, and Gaussian pulses are plotted in blue, green, and red, respectively. For each pulse shape, the color intensity varies from light to dark corresponding to $W/t^* = 0.0, 0.5, 1.0, 1.5, 2.0$.
}
\label{fig:fig3.3.2}
\end{figure}

As shown in  Fig.(\ref{fig:fig3.3.1}), A distribution function can be obtained after the interaction pulse with little dependence on the precise $T_{\text{ave}}$ value. From this distribution function, a fit of the Fermi-Dirac distribution near the Fermi energy allows us to obtain the effective final temperature or its inverse. Fig. (\ref{fig:fig3.3.2}) shows the inverse of the effective temperature as a function of $U_{\mathrm{max}}$ for different disorder strengths. This analysis is in general agreement with that of the change in total energy. The triangular pulse produces the smallest variation in temperature, indicating a more adiabatic response compared to the other pulse shapes. Moreover, stronger disorder promotes more adiabatic behavior across all pulse shapes.

%\section{Discussion}

%%%%%%%%%%%%%%%%%%%%%%%%%%%%%%%%%%%%%%%%%%
\section{Conclusions}
\label{sec:Conclusion}

We have used the nonequilibrium DMFT+CPA\cite{NeqDMFT_CPA1, NeqDMFT_CPABinaryDisorder, NeqDMFT_CPAThermalization, NeqDMFT_CPA2} solution to appropriately treat the nonequilibrium dynamics of an interaction disordered system across an interaction pulse, whereby the interaction starts off at zero and is increased to a maximum value before then being switched off again, according to a given pulse shape (rectangular, triangular, and Gaussian). We probe adiabaticity through the residual total energy in the system or the variation of the effective temperature after the interaction pulse. Through these solutions, we find that independently of the disorder strength and pulse shape, longer pulse durations reduce the change in total energy in the system, as would be generally expected. Most importantly, across the different pulse shapes considered, we find a robust negative correlation between disorder strength and the change in total energy across the interaction modulation. Namely, increasing the disorder strength systematically suppresses the change in total energy after the interaction is returned to zero, indicating a more adiabatic response of the system to the interaction pulse. These two effects, disorder-induced and duration-induced adiabaticity, are consistently observed across all three protocol shapes. Among the pulse shapes considered, the triangular protocol displays the smallest change in total energy in the system under comparable conditions, demonstrating the most adiabatic response. 
Although the different pulse protocols are chosen to have the same
total interaction area $\int U(t)\,dt$, they modify the interaction strength at different rates in time. In particular, the square pulse maintains the maximum interaction strength over an extended plateau, while the Gaussian pulse, although smooth, still spends a significant portion of the evolution near its peak value depending on the pulse width. By contrast, the triangular pulse reaches the maximum interaction only instantaneously and otherwise increases/decreases the interaction with a uniform rate between zero and the maximum value. Since the nonequilibrium response becomes stronger when the interaction strength is large, the longer time spent near $U_{\max}$ in the square and Gaussian protocols leads to larger energy absorption when the interaction is returned to zero. The triangular pulse therefore produces the smallest increase in the total energy and thus, the most adiabatic response.
This observation suggests that in addition to pulse duration, pulse protocols minimizing the time spent near the maximum interaction strength may provide a useful avenue for designing more optimal adiabatic driving schemes.
Altogether, our results identify disorder, protocol duration, and protocol shape as key factors governing both energy and temperature evolution during interaction modulation i.e. as control parameters for more adiabatic evolution.  

%The disorder strength $W$ shows a negative correlation with the change in total energy, indicating that the process becomes more adiabatic as $W$ increases. This trend is consistently observed across all pulse types (rectangular, triangular, and Gaussian). Similarly, the pulse period also exhibits a negative correlation with the change in total energy, demonstrating that longer periods enhance adiabaticity. This effect, too, is consistently present for all three pulse types. Consistent with these energy-based measures, the analysis of the effective final temperature shows that stronger disorder and smoother pulse protocols lead to smaller temperature changes after the pulse, providing further evidence that disorder and pulse shaping promote a more adiabatic thermal response of the system.

%%%%%%%%%%%%%%%%%%%%%%%%%%%%%%%%%%%%%%%%%%
%\section{Patents}

%This section is not mandatory, but may be added if there are patents resulting from the work reported in this manuscript.

%%%%%%%%%%%%%%%%%%%%%%%%%%%%%%%%%%%%%%%%%%
\vspace{6pt}

%%%%%%%%%%%%%%%%%%%%%%%%%%%%%%%%%%%%%%%%%%

%%%\acknowledgments{}

%%%%%%%%%%%%%%%%%%%%%%%%%%%%
%% Optional
% \appendixtitles{no} 
% \appendixstart
% \appendix
% \section[\appendixname~\thesection]{}
% \subsection[\appendixname~\thesubsection]{}
% The appendix is an optional section that can contain details and data supplemental to the main text---for example, explanations of experimental details that would disrupt the flow of the main text but nonetheless remain crucial to understanding and reproducing the research shown; figures of replicates for experiments of which representative data are shown in the main text can be added here if brief, or as Supplementary Data. Mathematical proofs of results not central to the paper can be added as an appendix.

% \section[\appendixname~\thesection]{}
% All appendix sections must be cited in the main text. In the appendices, Figures, Tables, etc. should be labeled, starting with ``A''---e.g., Figure A1, Figure A2, etc.

%%%%%%%%%%%%%%%%%%%%%%%%%%%%%%%%%%%%%%%%%%
%\isPreprints{}{% This command is only used for ``preprints''.
% ACS format

\begin{acknowledgments}
\noindent This work was supported by the U.S. Department of Energy, Office of Science, Basic Energy Sciences, under Award \#DE-SC0024139. We thank V. Oganesyan for useful discussions.
\end{acknowledgments}

\end{document}